\documentclass[fleqn,usenatbib]{mnras}

\usepackage[T1]{fontenc}
\usepackage{ae,aecompl}
\usepackage{bm}
\usepackage{graphicx}	% Including figure files
\usepackage{amsmath}	% Advanced maths commands
\usepackage{amssymb}	% Extra maths symbols

\usepackage{txfonts}

\title[Polar alignment of a protoplanetary disc with mass]{Polar alignment of a protoplanetary disc around an eccentric binary III: Effect of disc mass}

\author[Martin \& Lubow]{Rebecca G. Martin$^{1}$ and Stephen
  H. Lubow$^2$\\ $^{1}$Department of Physics and Astronomy, University
  of Nevada, Las Vegas, 4505 South Maryland Parkway, Las Vegas, NV
  89154, USA \\ $^2$Space Telescope Science Institute, 3700 San Martin
  Drive, Baltimore, MD 21218, USA}

\date{}
\pagerange{\pageref{firstpage}--\pageref{lastpage}} 
\pubyear{2019}

\begin{document}
\maketitle
\label{firstpage}

\begin{abstract}
\cite{Martin2017} found that an initially sufficiently misaligned low mass  protoplanetary disc around an eccentric binary undergoes damped nodal oscillations of tilt angle and longitude of ascending node.  Dissipation causes evolution towards  a stationary state of polar alignment in which the disc lies perpendicular to the binary orbital plane with angular momentum aligned to the eccentricity vector of the binary.  We use hydrodynamic simulations and analytic methods to investigate how the mass of the disc affects this process. The simulations suggest that a disc with nonzero mass settles into a stationary state in the frame of the binary, the generalised polar state, at somewhat lower levels
of misalignment with respect to the binary orbital plane, in agreement with the analytic model.
Provided that discs settle into this generalised polar state, the observational determination of the misalignment angle
and binary properties can be used to determine the mass of a circumbinary disc.
We apply this constraint to the circumbinary disc in HD 98800.
We obtain analytic criteria for polar alignment of a circumbinary ring with mass that approximately agree with the simulation results.
Very broad misaligned discs undergo breaking, but the inner regions at
least may still evolve to a polar state. The long term evolution of the disc depends on the evolution of the binary eccentricity that
we find tends to decrease. 
Although the range of parameters required for polar alignment decreases somewhat with increasing disc mass,
such alignment
appears possible for a broad set of initial conditions expected
in protostellar circumbinary discs. 
\end{abstract}

\begin{keywords} accretion, accretion discs -- binaries: general --
  hydrodynamics -- planets and satellites: formation
\end{keywords}

\section{Introduction}
\label{intro}

During the star formation process, misaligned discs around binary stars may be formed through chaotic accretion  \citep[e.g.][]{McKee2007,Bate2003,Monin2007,Bateetal2010, Bate2018}  or stellar flybys \citep{Clarke1993,Xianggruess2016,Cuello2019}. Observations of circumbinary discs suggest misalignments may be common \citep[e.g.][]{Winn2004,Chiang2004,Capelo2012,Brinch2016,Kennedy2012, Aly2018,Czekala2019}. The planet formation process in these discs will be altered by the torque from the binary that is not present in the single star case \citep[e.g.][]{Nelson2000,Mayer2005,Boss2006, Martinetal2014b,Fu2015,Fu2015b,Fu2017,Franchini2019}. Furthermore, giant planets that form in a misaligned disc may no longer remain coplanar to the disc \citep{Picogna2015,Lubow2016,Martin2016}. In order to understand the observed properties of exoplanets, we first need to explain the disc evolution in misaligned systems. 

A misaligned circumbinary disc around a {\it circular} orbit binary undergoes uniform
nodal precession with constant tilt. The angular momentum vector of the disc
precesses about the binary angular momentum vector. For a sufficiently warm and compact disc,  the disc precesses as a solid body \citep[e.g.,][]{LP1997}. Dissipation within the disc leads to alignment with the binary orbital plane
\citep{PT1995, Lubow2000, Nixonetal2011b,Nixon2012,Facchinietal2013,
  Lodato2013,Foucart2013,Foucart2014}.

Massless (test) particles that orbit around an eccentric orbit binary can undergo nodal libration oscillations of the tilt angle and the longitude of the ascending node, if the particle's orbital plane is sufficiently misaligned with the binary's orbital plane \citep[e.g.][]{Verrier2009,Farago2010,Doolin2011}. Rather than precessing about the angular momentum vector of the binary,
such particles instead precess about the eccentricity vector of the binary.

Recently, we found that a low mass warm protostellar circumbinary disc around an eccentric orbit binary can evolve towards polar (perpendicular) alignment with respect to the binary orbital plane for sufficiently high initial inclination \citep{Martin2017}. 
The  tilt evolution occurs due to damping of the libration oscillations by dissipation in the disc and the disc angular momentum aligns to the eccentricity vector of the binary. 
\cite{Aly2015} found that cool discs that orbit binary black hole systems can also undergo such oscillations and polar alignment. 
This mechanism operates for sufficiently large misalignment angle \citep{Aly2015, Lubow2018, Zanazzi2018}.

In \cite{Lubow2018} and \cite{Martin2018}, with analytic and numerical models, we extended the parameter space studied to include different disc properties such as viscosity, temperature, size and inclination and binary properties such as eccentricity and binary mass ratio.  For low initial inclination, a disc around an {\it eccentric} orbit binary undergoes tilt oscillations and non--uniform precession as it evolves towards alignment with the binary angular momentum vector \citep{Smallwood2019}.

For an orbiting particle with significant mass, the mass of this body has an important effect on the evolution of the system because it can affect the binary orbit, unlike in the low mass or test particle case. This regime has previously been explored analytically \citep{Lidov1976,Ferrer1994,Farago2010,Zanazzi2018}. 
They found that the stationary misalignment alignment angle (fixed point) between the binary and particle orbital planes is reduced below $90^\circ$ and depends on the binary eccentricity and the particle angular momentum. Although mass of a protostellar disc is  much less than the mass
of the binary, a circumbinary disc can extend to an outer radius
that is much greater than the binary separation.
Consequently, the angular momentum of a circumbinary disc can sometimes be significant compared to the binary angular momentum.
 
In this work, we explore for the first time the  polar evolution of a circumbinary disc with significant mass around an eccentric binary  by means of hydrodynamic simulations.  
We  compare those results to the results for circumbinary particles with mass that effectively represent a ring with mass. 
The behaviour of a circumbinary particle or ring with mass provides some insight
into the behavior of discs. However, the disc is an extended object that experiences larger torques at smaller radii, while
its angular momentum is typically dominated by material at larger radii.
In addition, for the case of a disc with significant mass, the binary orbital properties, such as its eccentricity and semi--major axis, evolve due to two effects: first the tidal interaction with the disc and second, the accretion of circumbinary disc material. 

In Section~\ref{disc} we explore the evolution of a misaligned circumbinary disc with significant disc mass by means of hydrodynamic simulations.  
In Section \ref{sec:polarana}
we consider a model of a ring with mass
and obtain analytic expressions for the (generalised) polar inclination and the requirements for evolution to the polar state.
We discuss the applications of our results in Section~\ref{discussion} and we draw conclusions in Section~\ref{conc}.

\begin{table*}
\centering
\caption{Parameters of the initial circumbinary disc set up for binary
  with total mass $M$ and separation $a$. The disc may be in a circulating (C) or librating (L) state.} 
\begin{tabular}{lccccclcc}
\hline
Name & Figure       & $M_{\rm d}/M$ & $i/^\circ$& $e_{\rm b}$& $R_{\rm out}/a_{\rm b}$ & C/L & Number of particles& Broken\\
\hline
\hline
%run5
run1  & \ref{mass} &    0.001&60 & 0.5 &5& {\rm L} & 300,000& No\\
%run13
run2 & \ref{mass} &    0.01&60 & 0.5 & 5& {\rm L} & 300,000& No\\
%run23
run3 & \ref{mass} &    0.02&60 & 0.5 & 5& {\rm L} & 300,000 & No\\
%run14
run4 & \ref{mass} &    0.05&60 & 0.5 & 5& {\rm L} & 300,000& No \\
\hline
%run27
run5 & \ref{fig:coplanar} &   0.05 & 0 & 0.5 &5&  -  & 300,000& No\\
%run35
run6 & \ref{fig:coplanar} &   0.01 & 0 & 0.5 & 5& -  & 300,000& No\\
%run36
run7 & \ref{fig:coplanar} &   0.001 & 0 & 0.5 &5&  - & 300,000& No \\
\hline
%run30
run8 & \ref{inc2} &    0.05 & 20 & 0.5 & 5& {\rm C} & 300,000& No \\
%run29
run9 & \ref{inc2} &    0.05 & 40 &0.5 & 5&  {\rm C} & 300,000& No \\
%run48
run10 & \ref{inc2}	   &    0.05 & 50 & 0.5 & 5& {\rm C } & 300,000 & No\\
%run28
run11 & \ref{inc2} &    0.05 & 80 & 0.5 & 5& {\rm L}  & 300,000& No\\
\hline
%run49
run12 & \ref{inc3} &   0.05 & 20 & 0.8&5& {\rm C}  & 300,000& No\\
%run55
run13 &             &   0.05 & 30 & 0.8& 5&{\rm C} & 300,000& No\\
%run50
run14 & \ref{inc3} &   0.05 & 40 & 0.8 &5& {\rm L} & 300,000& No\\
%run51
run15 &\ref{inc3}&    0.05 & 60 &  0.8& 5&{\rm L}& 300,000& No\\
%run52
run16 & \ref{inc3}&    0.05 & 80 &  0.8& 5& {\rm L}& 300,000& No\\
\hline 
%run61
run17 & \ref{inc4} & 0.001 & 60 & 0.5 & 10 & L& 600,000& No \\
%run59
run18 & \ref{inc4}& 0.01 & 60 & 0.5 & 10 & L& 600,000 & No\\
%run58
run19 & \ref{inc4}& 0.02 & 60 & 0.5 & 10 & L & 600,000& No \\
%run56b
run20 & \ref{inc4}& 0.05 & 60 & 0.5 & 10 & L & 600,000& No\\
\hline
%run60
run21 & \ref{inc5} & 0.05 & 60 & 0.5 & 20 & L & 600,000 & Yes\\
\hline
\end{tabular}
\label{tab}
\end{table*} 

\section{Circumbinary disc Simulations}
\label{disc}

In this section we explore the evolution misaligned circumbinary disc with significant mass around a binary star system.  We apply the smoothed particle hydrodynamics (SPH; e.g. \citealt{Price2012a,Price2007}) code {\sc phantom} \citep{PF2010,LP2010,Price2018} that has been used extensively for simulations of misaligned accretion discs \citep[e.g.][]{Nixon2012,Nixonetal2013,Martinetal2014b,Fu2015}.

\subsection{Simulation set--up}

\begin{figure}
\centering
\includegraphics[width=\columnwidth]{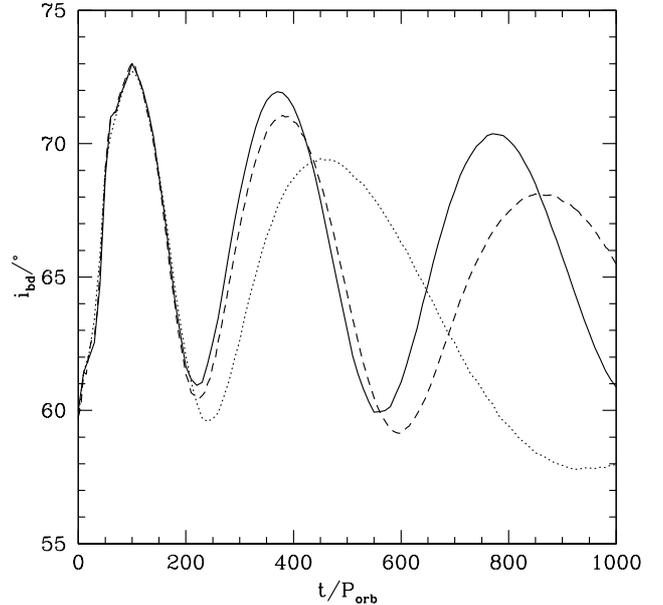}
\caption{  Resolution study for run4. The inclination of the disc relative to the binary as a function of time for the high mass simulation with $M_{\rm d}=0.05\,M$, initial inclination $i=60^\circ$, $e_{\rm b}=0.5$, $R_{\rm in}=2\,a_{\rm b}$ and $R_{\rm out}=5\,a_{\rm b}$. The model with the solid line initially has $1 \times 10^6$ particles, the dashed line has $3 \times 10^5$, and the dotted line has $1 \times 10^5$. The inclinations are measured at disc radius $R=5\,a_{\rm b}$.} 
\label{res}
\end{figure}

\begin{figure*}
\centering
\includegraphics[width=\columnwidth]{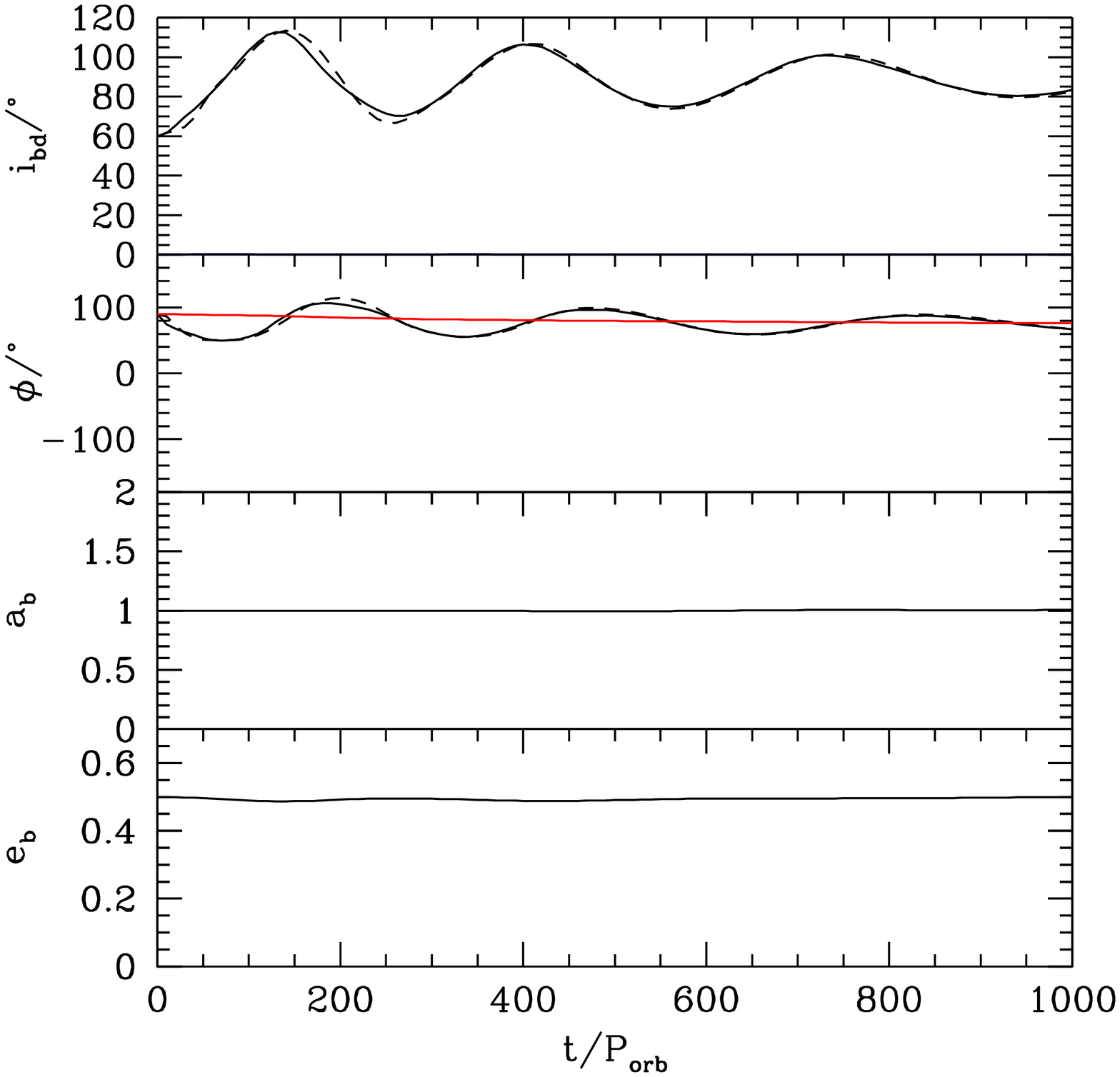}
\includegraphics[width=\columnwidth]{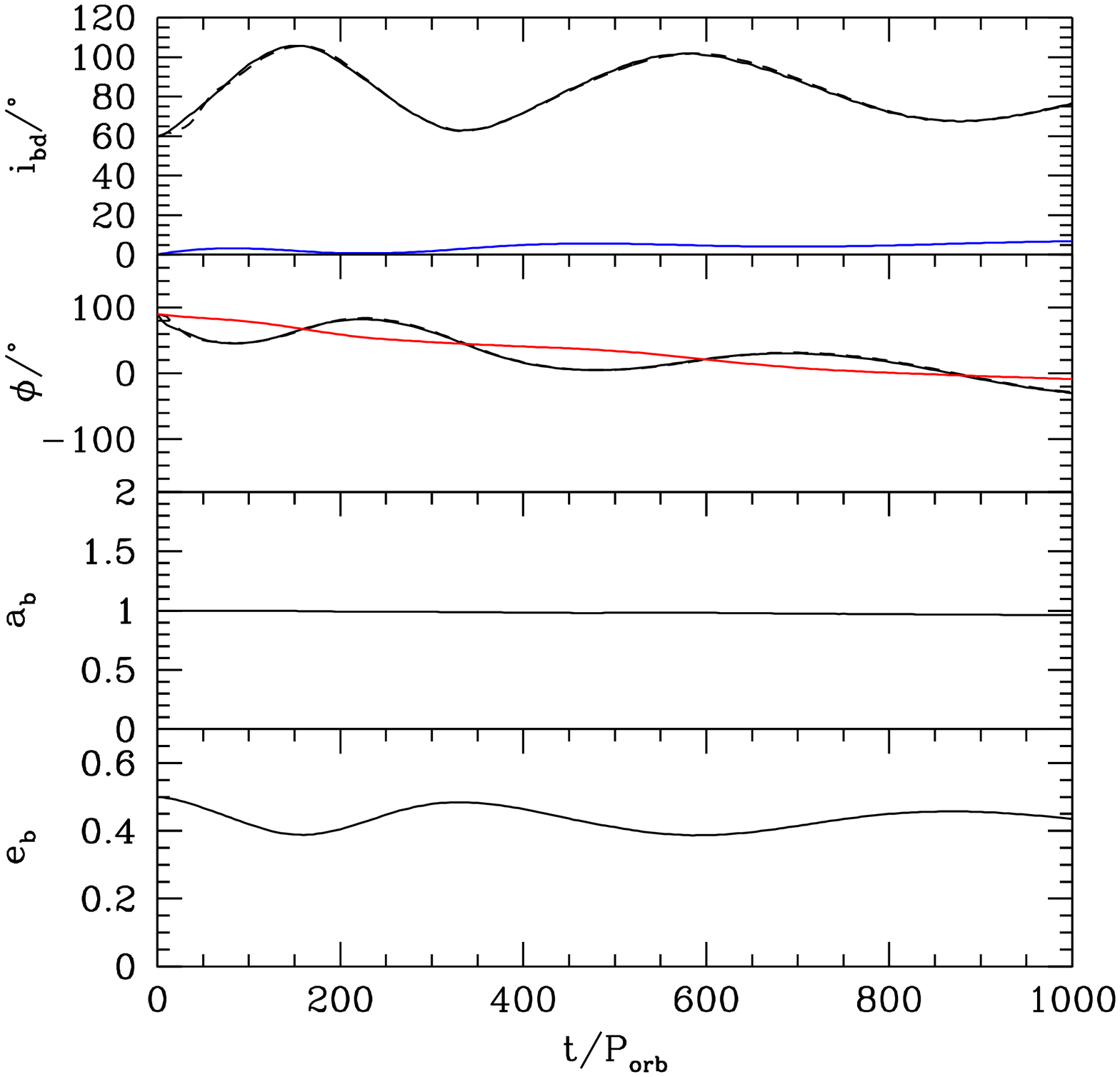}
\includegraphics[width=\columnwidth]{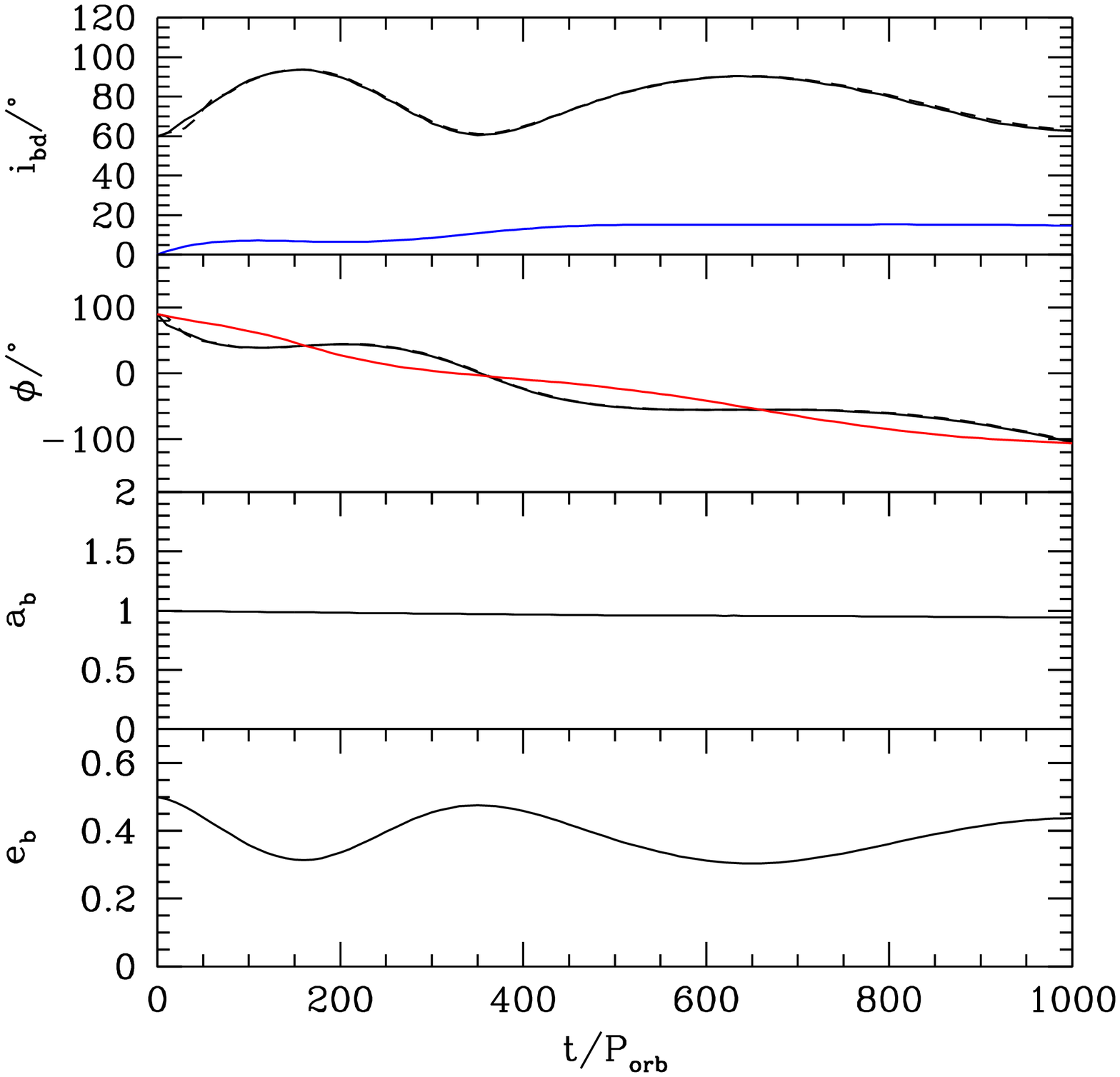}
\includegraphics[width=\columnwidth]{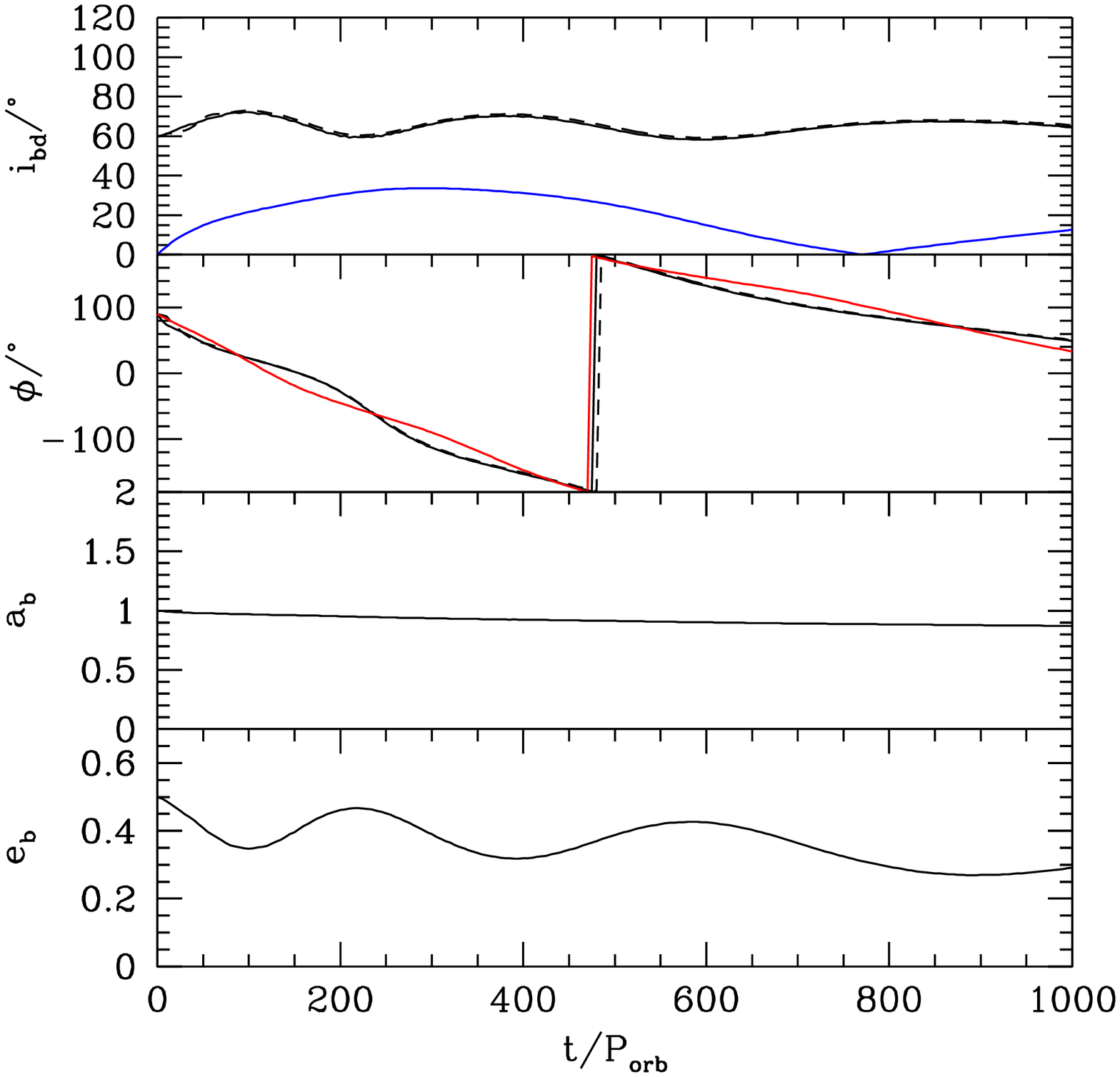}
\caption{Simulations of a circumbinary disc around an equal mass binary with initial binary eccentricity $e_{\rm b}=0.5$, initial disc inclination $i=60^\circ$ and $H/R=0.1$ at the disc inner edge of $R_{\rm in}=2\,a_{\rm b}$. The disc outer edge is initially at $R_{\rm out}=5\,a_{\rm b}$. In the upper two panels, the solid lines are for a radius of $R=3\, a_{\rm b}$ and the dashed lines for $R=5\,a_{\rm b}$.  Top left: $M_{\rm d}=0.001\,M$ (run1). Top right: $M_{\rm d}=0.01\,M$ (run2). Bottom left: $M_{\rm d}=0.02\,M$  (run3). Bottom right: $M_{\rm d}=0.05\,M$ (run4). Upper panels: inclination of the disc angular momentum vector relative to the binary angular momentum vector, $i_{\rm bd}$ (Equation (\ref{ibd})). The blue lines plot the inclination of angular momentum vector of the binary relative to the reference plane, $i_{\rm b}$ (Equation (\ref{ib})). Second panels: precession angles $\phi$. The black lines show the nodal precession angle for the disc $\phi_{\rm d}$ (Equation~\ref{phid}). The red lines show the binary eccentricity vector phase angle $\phi_{\rm b}$ (Equation (\ref{phib})). Third panels: semi--major axis of the binary $a_{\rm b}$. Lower panels: magnitude of the eccentricity of the binary, $e_{\rm b}$.   }
\label{mass}
\end{figure*}

Table~\ref{tab} summarises the parameters and some results for all of the simulations that we describe in this section. The binary has components with masses $M_1=M_2=0.5\,M$, where the total mass is $M=M_1+M_2$. The binary orbits with semi--major axis $a_{\rm b}$ and eccentricity vector $\bm{e_{\rm b}}=(e_{x\rm b},e_{y\rm b},e_{z\rm b})$. The binary orbit is initially in the $x-y$ plane with eccentricity vector $\bm{e_{\rm b}}=(1,0,0)$.  This $x-y$ plane serves as a reference plane for the orbital elements described below.
The binary begins at apastron separation.

Initially the circumbinary disc is  misaligned  to the binary orbital plane by inclination angle $i$.   The surface density is initially distributed by a power law $\Sigma \propto R^{-3/2}$ between the initial inner radius $R_{\rm in}=2\,a_{\rm b}$ up to the initial outer radius $R_{\rm out}$. Typically we take $R_{\rm out}=5\,a_{\rm b}$, but we do consider some larger values also. The initial inner disc truncation radius is chosen to be that of a tidally truncated coplanar disc \citep{Artymowicz1994}. However, the disc spreads both inwards and outwards during the simulation. As described in \cite{Lubow2018}, the inner edge of the disc extends closer to the binary because the usual gap-opening Lindblad resonances are much weaker on a polar disc around an eccentric binary than in a coplanar disc around a circular or eccentric orbit binary \citep[see also ][]{Lubowetal2015,Nixon2015,Miranda2015}. We take the \cite{SS1973} $\alpha$ parameter to be 0.01 in our simulations. The disc viscosity is implemented in the usual manner by adapting the SPH artificial viscosity according to \cite{LP2010}. The disc is locally isothermal with sound speed $c_{\rm s}\propto R^{-3/4}$ and the disc aspect ratio varies with radius as  $H/R\propto R^{-1/4}$. Hence, $\alpha$ and the smoothing length $\left<h \right>/H$ are constant over the radial extent of the disc \citep{LP2007}. We take $H/R=0.1$ at  $R_{\rm in}=2\,a_{\rm b}$. We examined the effects of these two parameters in \cite{Lubow2018}.  Particles in the simulation are removed if they pass inside the accretion radius for each component of the binary at $0.25\,a_{\rm b}$.  

We ignore the effects of self--gravity in our calculations. 
Self-gravity can play an important role in cases where apsidal precession
plays an important role in eccentric discs, as occurs for Kozai-Lidov discs
\citep{Batygin2011, Fu2017}. However, for an initially circular disc, as we assume,
we find that the discs remain quite circular, as is expected since eccentricity is a constant of motion for ballistic circumbinary particles (in the quadrupole approximation) \citep{Farago2010}.
Instead, nodal precession plays the key role in the dynamics of circumbinary discs. But self-gravity has no effect on the nodal precession rate of a flat disc.
Provided that the disc is flat, the stationary tilt condition that we describe in Section \ref{sec:stationary} should be independent of self--gravity. 

Self--gravity could have some influence on the level of warping and that in turn could affect
the tilt evolution time.
Narrow discs are more likely to be affected by self--gravity for a fixed disc mass, since the surface density is higher. However, the initial Toomre parameter is $Q>2.3$ over the radial extent of the disc for the narrowest ($R_{\rm out}=5\,\rm a_{\rm b}$), highest disc mass ($M_{\rm d}=0.05\,M$) that we consider. The value of $Q$ increases over time in our simulations.
Discs of larger radial extent that warp or break may be affected by self-gravity.  However, the wider discs we consider have larger initial Toomre parameter $Q>3.1$ for initial disc outer radius of $20\,a_{\rm b}$, and in the regions that warping or breaking takes place, $Q \gtrsim 10$. Thus, self--gravity is not important in our calculations.

In order to present a large number of simulations in this work, we choose to use $3 \times 10^5$ particles in most of our following simulations. We found this number to provide sufficient resolution in Figure~1 in \cite{Martin2018} for a time of about $1000\,P_{\rm orb}$, where $P_{\rm orb}$ is the orbital period of the binary.  Fig.~\ref{res}  shows the results of a similar resolution study, but for a disc with more mass. In the previous work, the convergence test started with a disc mass of $0.001\,M$, while in this study we use run4 of Table~\ref{tab} in which the initial disc mass is $0.05\,M$.  We find that the properties based on the first two oscillations are fairly well converged, based on the simulations with $3 \times 10^5$ and $1 \times 10^6$ particles. Lower resolution at late times leads to increased viscosity and more damping in the oscillations. But the behaviour in the two cases is quite similar in that the oscillations are centered about a binary--disc inclination  $i_{\rm bd} \sim 65^\circ$, rather than almost $90^\circ$ found for the lower mass disc in the earlier paper.

In order to provide adequate vertical resolution for a disc, we generally require that the smoothing length $h$ be less than the disc scale height $H$  \citep[e.g.,][]{Armitage1996}. 
The disc is resolved with  initial shell-averaged smoothing length per scale height $\left<h\right> /H \approx 0.25$ for $R_{\rm out}=5\,a_{\rm b}$. For simulations with larger initial disc outer radii $R_{\rm out}=10\,a_{\rm b}$ and $R_{\rm out}=20\,a_{\rm b}$, we use $6 \times 10^5$ particles initially and the disc is initially resolved with $\left<h\right> /H \approx 0.26$ and $\left<h\right> /H \approx 0.31$, respectively.  The value of $\left<h\right> /H$ does not change significantly over the disc over the times we simulate,  as we show later. 

A limitation of our simulations is that the flow in the central
gap region is not well resolved by the SPH code in these intrinsically 3D flows. The flow in that region takes the form of
 rapid low density gas streams \citep[e.g.,][]{Artymowicz1996, Munoz2019, Mosta2019}. This limitation introduces some uncertainty in the binary evolution.

In order to describe the evolution of the system, we compute the inclination of the disc relative to the instantaneous binary angular momentum as
\begin{equation}
i_{\rm bd}=\cos^{-1}(\bm{ l_{\rm b}}\cdot \bm{ l_{\rm d}}),
\label{ibd}
\end{equation}
where ${\bm l_{\rm b}}=(l_{x\rm b},l_{y\rm b},l_{z\rm b})$ is the unit vector in the direction of the binary angular momentum and $\bm{l_{\rm d}}=(l_{x\rm d},l_{y\rm d},l_{z\rm d})$ is a unit vector in the direction of the disc angular momentum vector. The longitude of ascending node phase angle for the disc is
\begin{equation}
\phi_{\rm d}=\tan^{-1}\left( \frac{l_{y\rm d}}{l_{x\rm d}} \right) +\frac{\pi}{2}.
\label{phid}
\end{equation}
We also determine the  phase angle of the eccentricity vector of the binary projected onto the reference plane. We define this phase angle as 
\begin{equation}
\phi_{\rm b}=\tan^{-1}\left( \frac{e_{y\rm b}}{e_{x\rm b}} \right)+\frac{\pi}{2}.
\label{phib}
\end{equation}
This phase is plotted as red lines in the figures that we describe later. The inclination of the binary relative to the reference plane varies in time and is defined as
\begin{equation}
i_{\rm b}=\cos^{-1}\left( l_{z\rm b} \right).
\label{ib}
\end{equation}
This angle is plotted as blue lines in the figures that we describe later.

\subsection{Effect of the disc mass on the disc alignment}

\begin{figure*}
\centering
%run1 (run5b)
\includegraphics[width=\columnwidth]{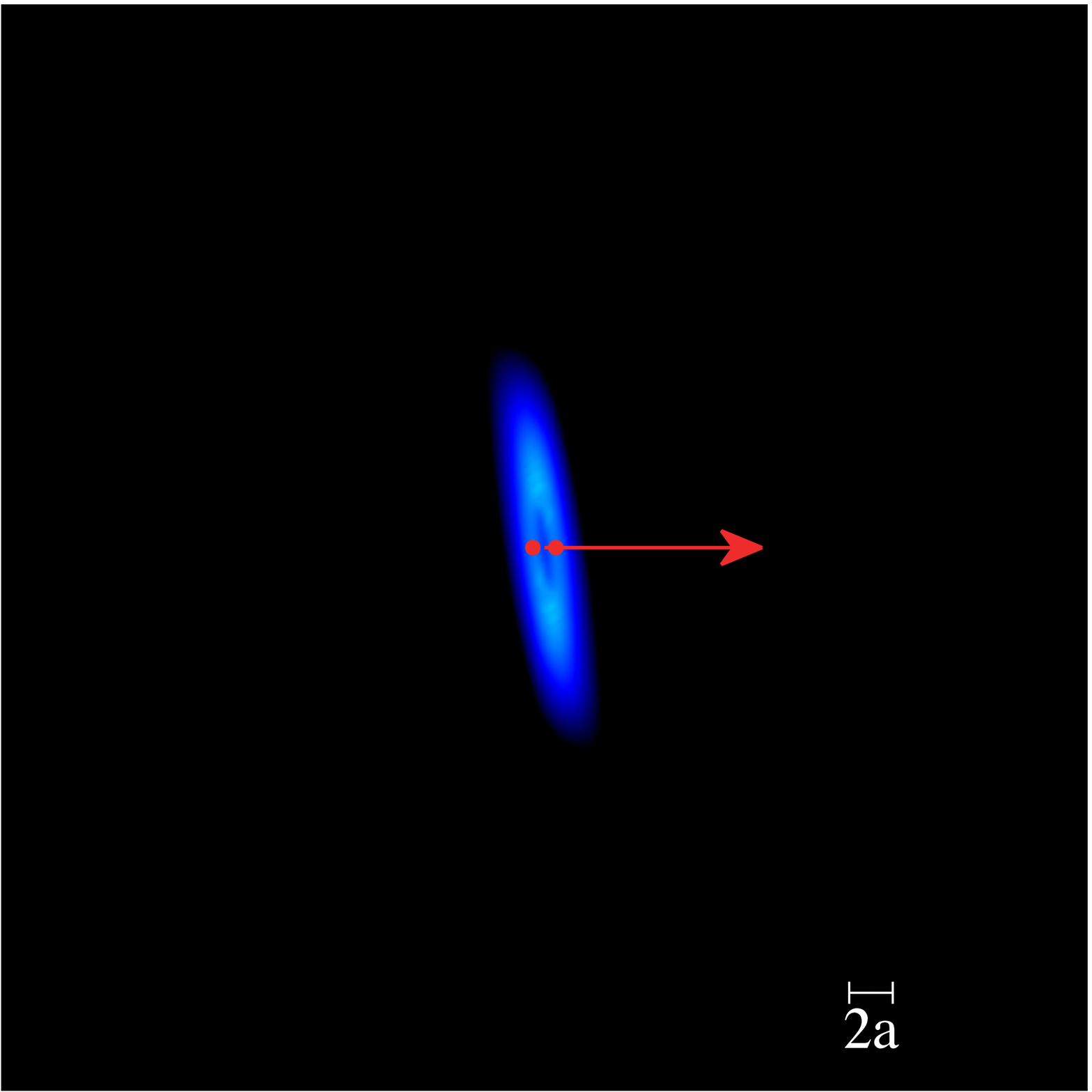}
%run4 (run14b)
\includegraphics[width=\columnwidth]{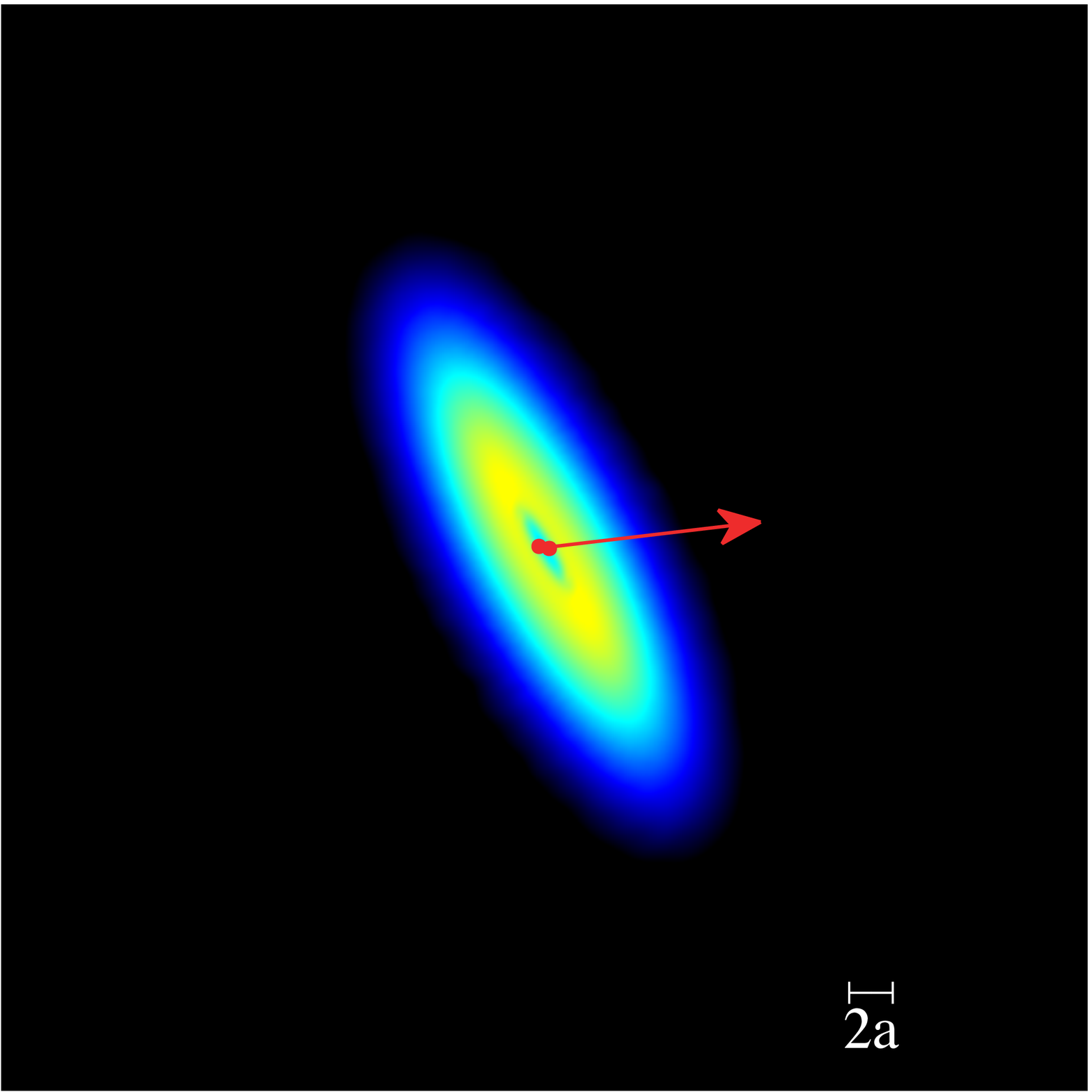}
%run56b
\includegraphics[width=\columnwidth]{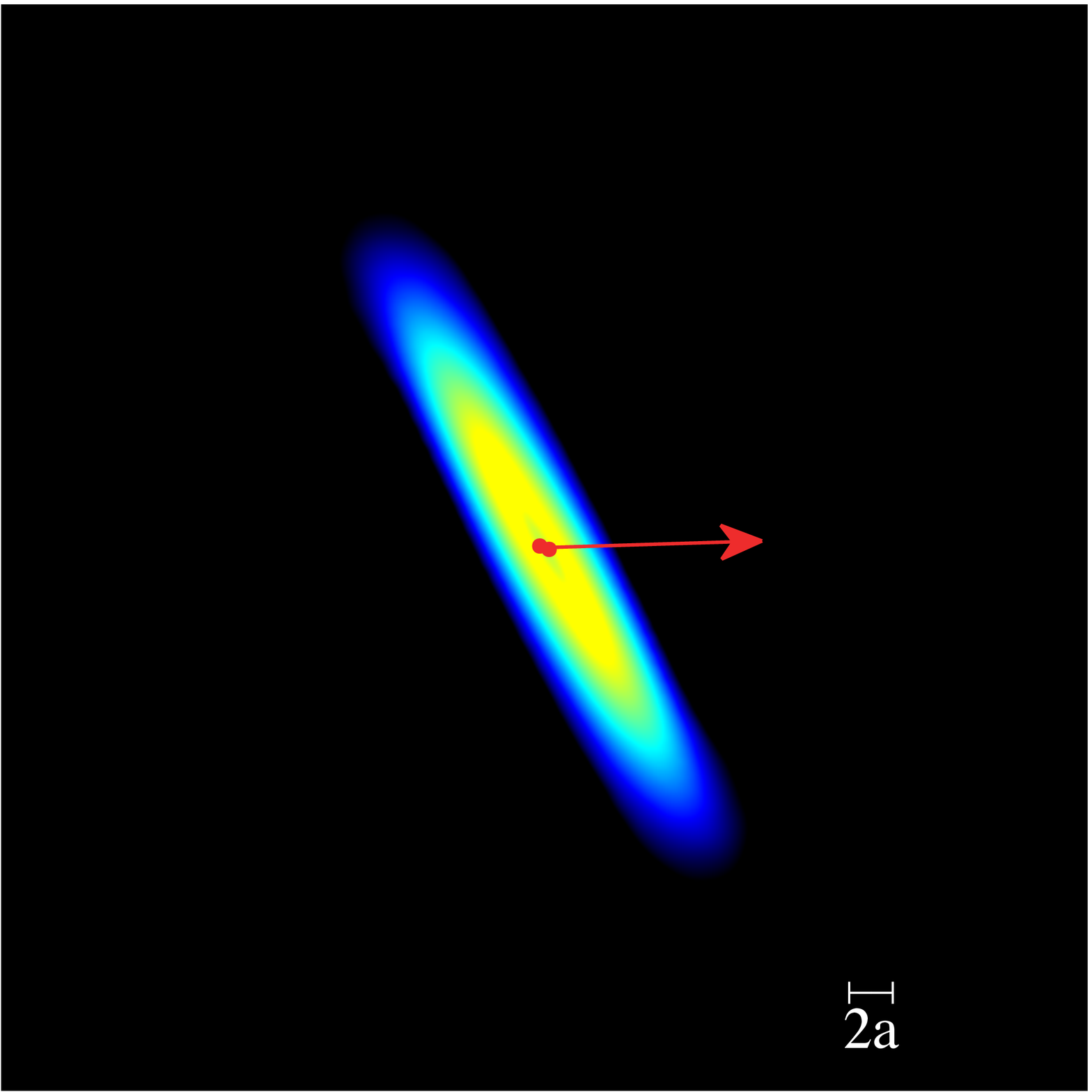}
%run21 (run60)
\includegraphics[width=\columnwidth]{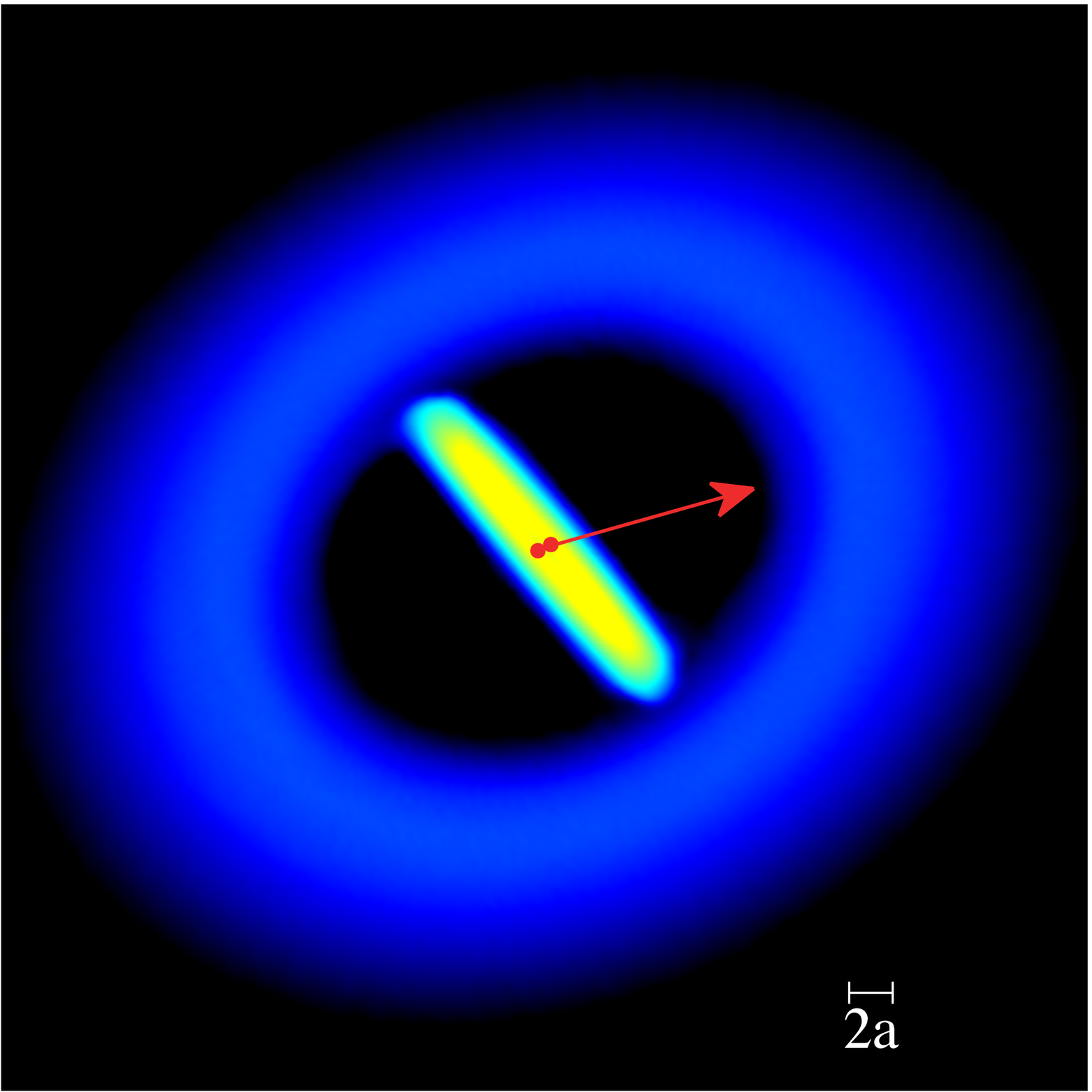}
\caption{  Circumbinary disc around an equal mass binary with initial binary eccentricity $e_{\rm b}=0.5$, initial disc inclination $i=60^\circ$ and $H/R=0.1$ at the initial disc inner edge of $R_{\rm in}=2\,a_{\rm b}$. The system is shown at a time of $t=1000\,P_{\rm orb}$.
Top left: The low mass narrow disc case $M_{\rm d}=0.001\,M$ and $R_{\rm out}=5\,a_{\rm b}$ (run1). Top right: The high mass narrow disc with $M_{\rm d}=0.05\,M$ and $R_{\rm out}=5\,a_{\rm b}$ (run4). Bottom left: The extended disc with $M_{\rm d}=0.05\,M$ and $R_{\rm out}=10\,a_{\rm b}$ (run20). Bottom right: The extended disc with $M_{\rm d}=0.05\,M$ and $R_{\rm out}=20\,a_{\rm b}$ (run21).  The $z$ axis corresponds to the initial binary angular momentum vector.  The viewing angle is rotated about the $z$ axis so that the binary eccentricity vector lies in the plane of the figure and its direction is shown by the red arrow. 
The red circles show the binary components. The colour denotes the gas density with yellow being about 3 orders of magnitude higher than blue.} 
\label{splash}
\end{figure*}

We first consider the effect of the disc mass on the standard disc model parameters shown in run1 of Table 1 which is the same model presented in \cite{Martin2017}. The binary is equal mass with an initial orbital eccentricity of 0.5.  The disc is initially inclined by $60^\circ$ to
the binary orbital plane.  We calculate disc properties by dividing the disc into 100 bins in spherical radius. Within each bin, we calculate the mean properties of the particles, such as the surface density, inclination, longitude of ascending node, eccentricity. 

The top left panel of Fig.~\ref{mass} shows the evolution of the disc with our standard parameters in run1.  We plot the evolution  at a disc radius of $r=3\,a_{\rm b}$ (solid lines) and $r=5\,a_{\rm b}$ (dashed lines). The disc acts like a solid body since these lines nearly overlap. As described in \cite{Martin2017}, the disc undergoes nodal libration in which the tilt and longitude of the ascending node oscillate. Dissipation causes the disc to evolve towards polar alignment where $i_{\rm bd}\approx 90^\circ$. The disc angular momentum vector aligns with the eccentricity vector of the binary and the disc approaches a nonprecessing state. For this low mass disc, there is little evolution of the binary separation, eccentricity vector (as shown by the red line), or inclination (as shown by the blue line).  The top left panel of Fig.~\ref{splash} shows the disc at a time of $t=1000\,\rm P_{\rm orb}$. The disc is close to polar alignment with the angular momentum of the disc being close to alignment with the binary eccentricity vector (shown in red).

The other panels in Fig.~\ref{mass} show the disc evolution with a higher initial mass of $M_{\rm  d}=0.01\,M$ (top right, run2), $M_{\rm d}=0.02\,M$ (bottom left, run3), and $M_{\rm d}=0.05\,M$ (bottom right, run4). Now the effect of the disc on the binary is no longer negligible. The binary undergoes apsidal precession (as seen by the red line in the phase angle plot), the binary inclination changes (see the blue lines), and the magnitude of the eccentricity of the binary oscillates and decays.

For all four disc masses in Fig.~\ref{mass},  the binary and disc phase angles are nearly the equal. The phase difference undergoes a small amplitude oscillation. 
 Over this time, the disc is then nodally librating with respect to the binary, rather than circulating. The libration indicates that the system is
 in a state where it lies above the critical level of misalignment
 for polar-like behavior. This 
 suggests that the system is undergoing evolution towards a  polar-like state as found in the low mass disc case. 
 There is a small reduction in binary semi-major axis that becomes larger
 with disc mass.
 In addition,  the binary eccentricity has declined somewhat after about $1000$ binary orbits.  The eccentricity oscillates, but the eccentricity
 decreases overall with disc mass.
The reduction of binary eccentricity suggests that over longer timescales the disc might eventually become coplanar with the binary, since the polar disc mechanism requires a certain level binary eccentricity.  Better resolution
 is required to study the longer term evolution.

\begin{figure}
\centering
\includegraphics[width=\columnwidth]{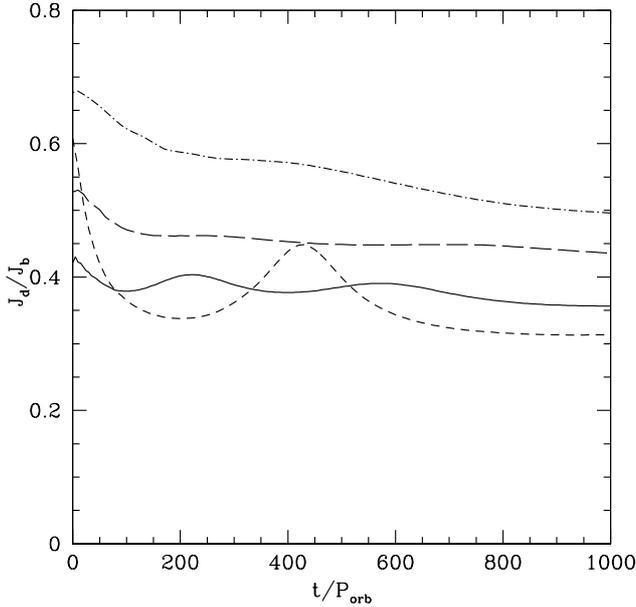}
\caption{The evolution of the ratio of the angular momentum of the disc to the angular momentum of the binary for run4 (solid line), run14 (short--dashed line), run20 (long--dashed line) and run21 (dot--dashed line).  }
\label{angmom}
\end{figure}

 Fig.~\ref{angmom} plots the evolution of the ratio of the angular momentum of the disc to that of the binary, $J_{\rm d}/J_{\rm b}$ for four different simulationss that all have $M_{\rm d}=0.05 M$, including the case plotted
 in the lower right panel of Fig.~\ref{mass} in the solid line.  The ratios oscillate in time because the eccentricity of the binary oscillates.
 In all four cases, the disc angular momentum
 is quite significant with 
 $J_{\rm d} \ga 0.3 J_{\rm b}$. 

 Unlike the very low mass disc case, a disc with significant mass evolves towards a highly misaligned nonprecessing state relative to the binary which is not perpendicular to the binary orbital plane. We define the stationary inclination angle that the disc is evolving towards as $i_{\rm bd}=i_{\rm s}$ where the disc precession rate  relative to the binary vanishes, i.e., the disc phase angle is stationary (denoted by subscript s) relative to the binary phase angle.  Only in the massless circumbinary disc case does the disc evolve to exactly polar alignment with $i_{\rm s}= 90^\circ$.

As we discuss in Section \ref{sec:stationary}, angle $i_{\rm s}$ decreases with increasing particle angular momentum.  Consequently, a narrow ring with the same orbital radius as the particle should also experience a decrease
in the $i_{\rm s}$ with increasing ring mass. 
Similar effects are expected for a disc. For the disc mass of $0.05\,M$ (bottom right panel of Fig. \ref{mass}), the binary-disc inclination oscillations are damping and
the disc is evolving towards $i_{\rm bd} = i_{\rm s}\approx 65^\circ$.
Thus, the mass of the disc plays an important role in the stationary orientation of the system. We discuss this further in Section~\ref{discussion}.  The top right panel of Fig.~\ref{splash} shows the high mass disc at a time of $t=1000\,P_{\rm orb}$. The disc is close to a stationary state in the frame of the binary, the generalised polar state, that has  a lower level of misalignment than the low mass disc (shown in the top left panel). 

\begin{figure}
\centering
\includegraphics[width=\columnwidth]{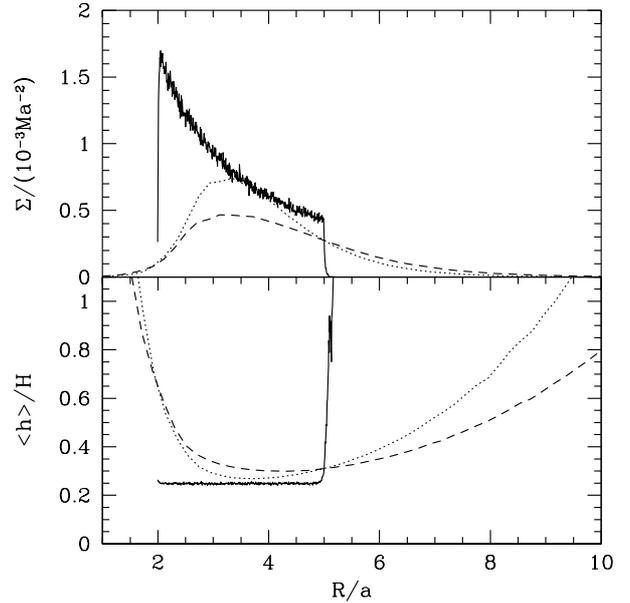}
\caption{  The surface density (upper) and smoothing length--to--disc scale height ratio (lower) as a function of radius in the disc for run4 at times $t=0$ (solid lines), 500 (dotted lines) and $1000\,P_{\rm orb}$ (dashed lines).} 
\label{smooth}
\end{figure}

 Fig.~\ref{smooth} shows the surface density and the smoothing length as a function of radius at three different times for run4, the high mass disc case. The disc is initially truncated at $R=5\,a$, but spreads outwards during the simulation. The disc is well resolved ($\left<h\right>/H<1$) out to $R=10\,a$ for the duration of the simulation except in regions of very low density, the innermost and outermost parts of the disc.

\subsection{Binary orbital evolution}
\label{coplanar}

\begin{figure}
\includegraphics[width=\columnwidth]{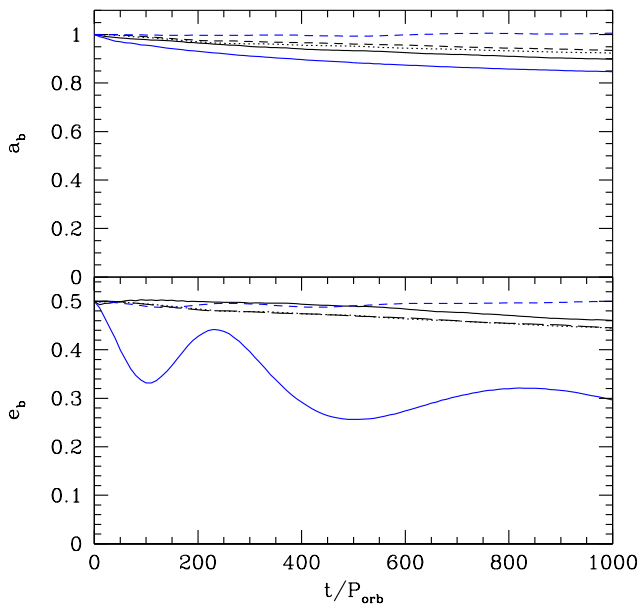}
\caption{The binary semimajor axis and eccentricity evolution due to a coplanar circumbinary disc. 
  The  initial disc mass is $0.05\,M$ (solid black lines, run5), $0.01\,M$ (dotted black lines, run6)
  and $0.001\,M$ (dashed black lines, run7) and otherwise the same initial disc properties. The blue lines show for comparison
  the standard parameters with a disc mass of $0.001\,M$ and an
  inclination of $60^\circ$ (dashed blue lines, run1)  and a disc mass of $0.05\,M$ and an
  inclination of $60^\circ$ (solid blue lines, run4)}.
\label{fig:coplanar}
\end{figure}

The binary orbital evolution is affected by a disc with significant mass.
The evolution of the binary angular momentum is determined by both the
accretion of angular momentum from the disc and the gravitational
torques from the disc.
 The former leads to an accretional torque. Due to our limited resolution in the inner gap, the effects of this torque on the binary evolution are somewhat uncertain.
The effects of the accretional torque on the binary have been difficult to determine even in 2D simulations \cite[e.g.,][]{Munoz2019}.

The gravitational torque contributions to binary orbit changes involve the interaction
of the disc with binary resonances that in turn depend on properties
of the binary.  In the coplanar binary-disc case for small binary eccentricity,
the theory of resonant disc gravitational torques suggests that the binary eccentricity  increases due to the dominant effects of a single resonance in the disc \citep{Artymowicz1991}. However, at higher binary eccentricities many resonances can lie within the disc, some of which cause binary eccentricity damping. For eccentricities $e_{\rm b} \sim 0.5$ or greater, the binary  eccentricity
growth rate due to disc resonances may become very small or even become negative \citep{Lubow1992a}.

Simulations by \cite{Artymowicz1991} found that for an initially low
eccentricity binary, $e_{\rm b}=0.1$, the eccentricity of the binary increased due to
gravitational interactions with the disc.  \cite{AN2005} confirmed
this increase in eccentricity along with a decreasing
semi--major axis and suggested that this may solve the final parsec
problem of merging massive black hole binaries, at least for extreme
mass ratio binaries. More recently, \cite{Shi2012} performed the first
3D magnetohydrodynamic (MHD) simulations of a circumbinary disc around
an equal mass circular binary. They found that the MHD stresses
allowed accretion on to the binary resulting in the semi-major
axis increasing slowly. \cite{Miranda2017} and \cite{Munoz2019} performed hydrodynamical
simulations with a grid code for a range of binary eccentricities and
found that the binary separation increases in time.

Because of the sensitivity of the binary evolution to system
parameters, we consider here for comparison the evolution in our SPH
models in the coplanar case. The black lines in Fig.~\ref{fig:coplanar} show the evolution of the binary 
in three coplanar disc simulations for varying disc mass. The eccentricity of the binary decreases in time while the semi-major axis also
decreases. The eccentricity change is somewhat insensitive to the mass of the
disc while the semi-major axis decreases more quickly for larger disc
mass. For comparison, in the blue lines in Fig.~\ref{fig:coplanar} we also show the binary orbit
evolution in our standard inclination parameters of run1 and the high mass disc of run4. A low mass inclined disc leads to very little binary orbital evolution over the timescale of our simulation.  The higher mass disc leads to more binary eccentricity evolution, as we  discussed in the previous subsection.

\subsection{Critical inclination for circulating and librating solutions}

There is a critical inclination above which the disc is librating and below which it circulates. In \cite{Lubow2018} we found that for a low mass disc, the critical inclination is close to that predicted for a test particle orbit.  As we discuss in Section \ref{sec:citinc}, the critical inclination for a third body with nonzero mass depends upon the angular momentum of the body. Here we consider the critical inclination for two different binary eccentricities. 

\subsubsection{Initial binary eccentricity $e=0.5$}
\label{eb0p5}

\begin{figure*}
\centering
\includegraphics[width=\columnwidth]{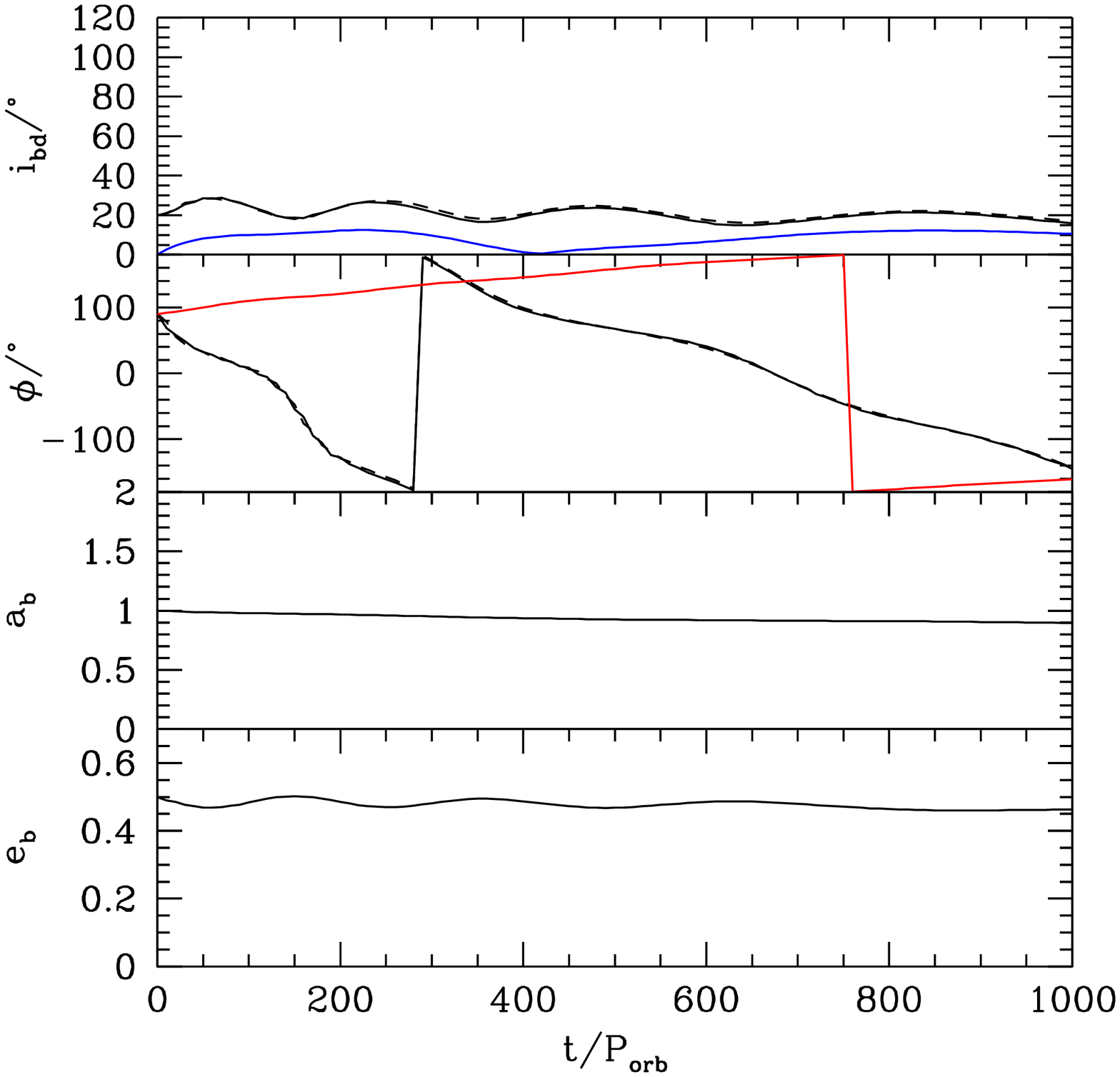}
\includegraphics[width=\columnwidth]{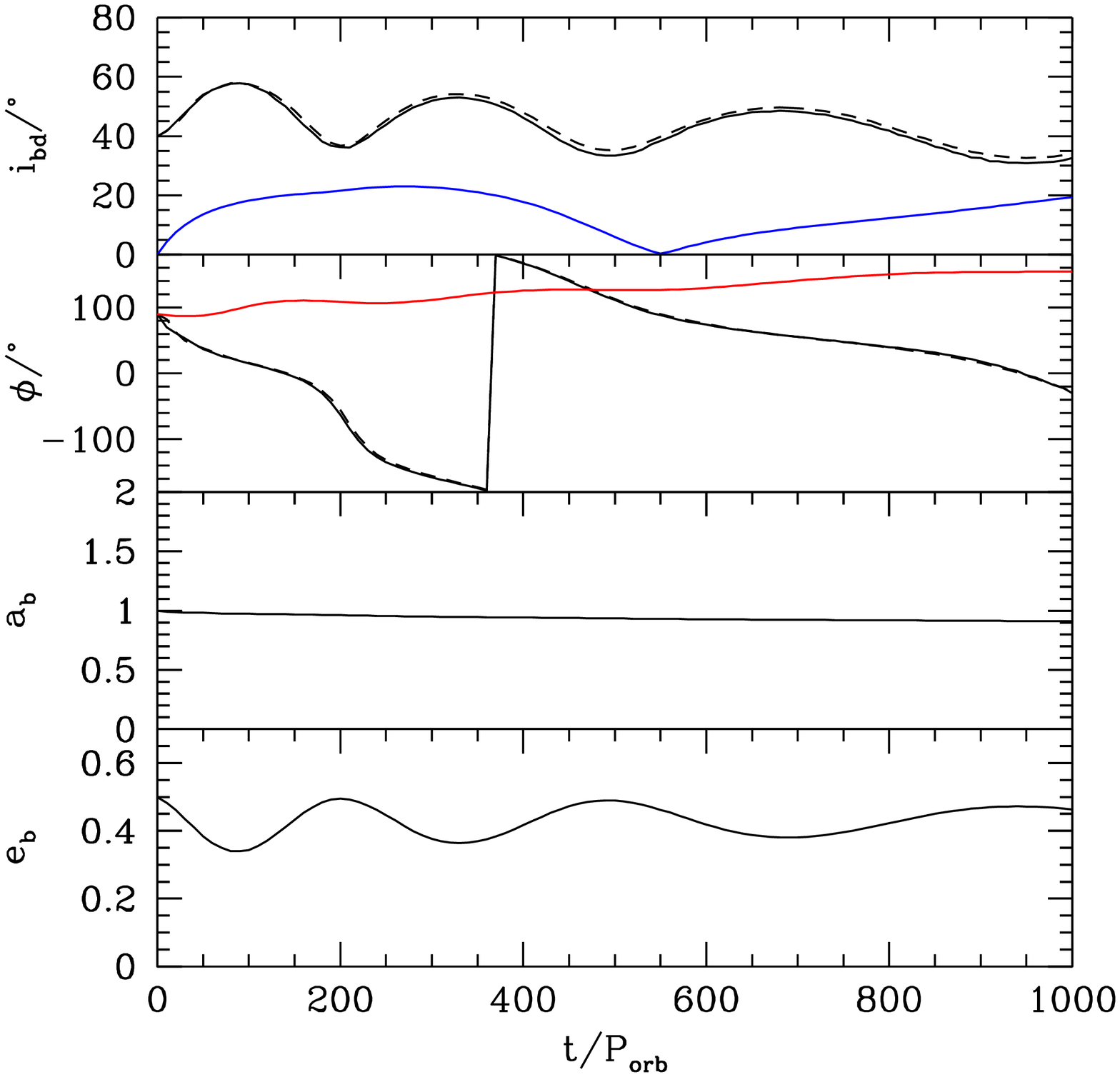}
\includegraphics[width=\columnwidth]{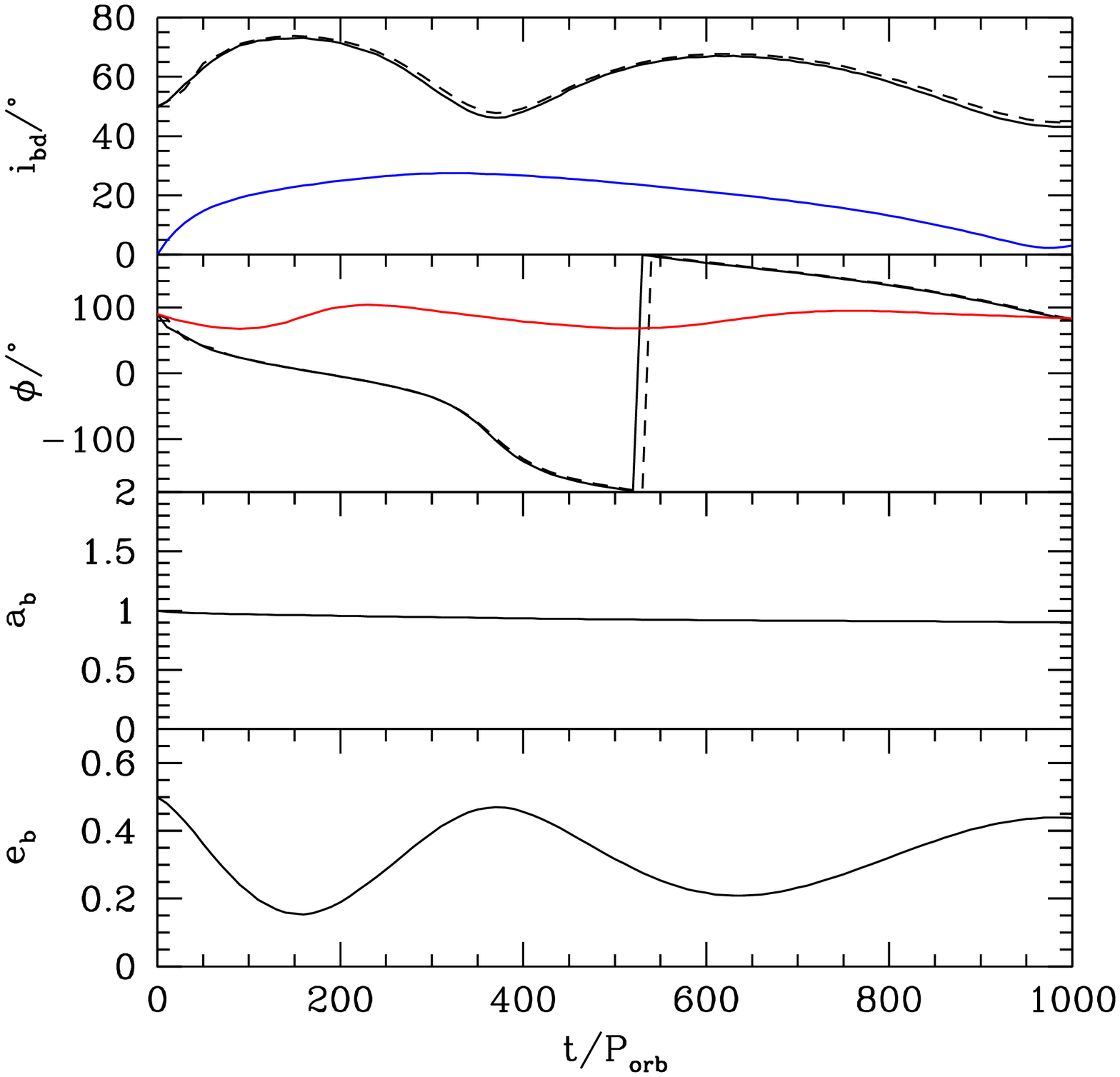}
\includegraphics[width=\columnwidth]{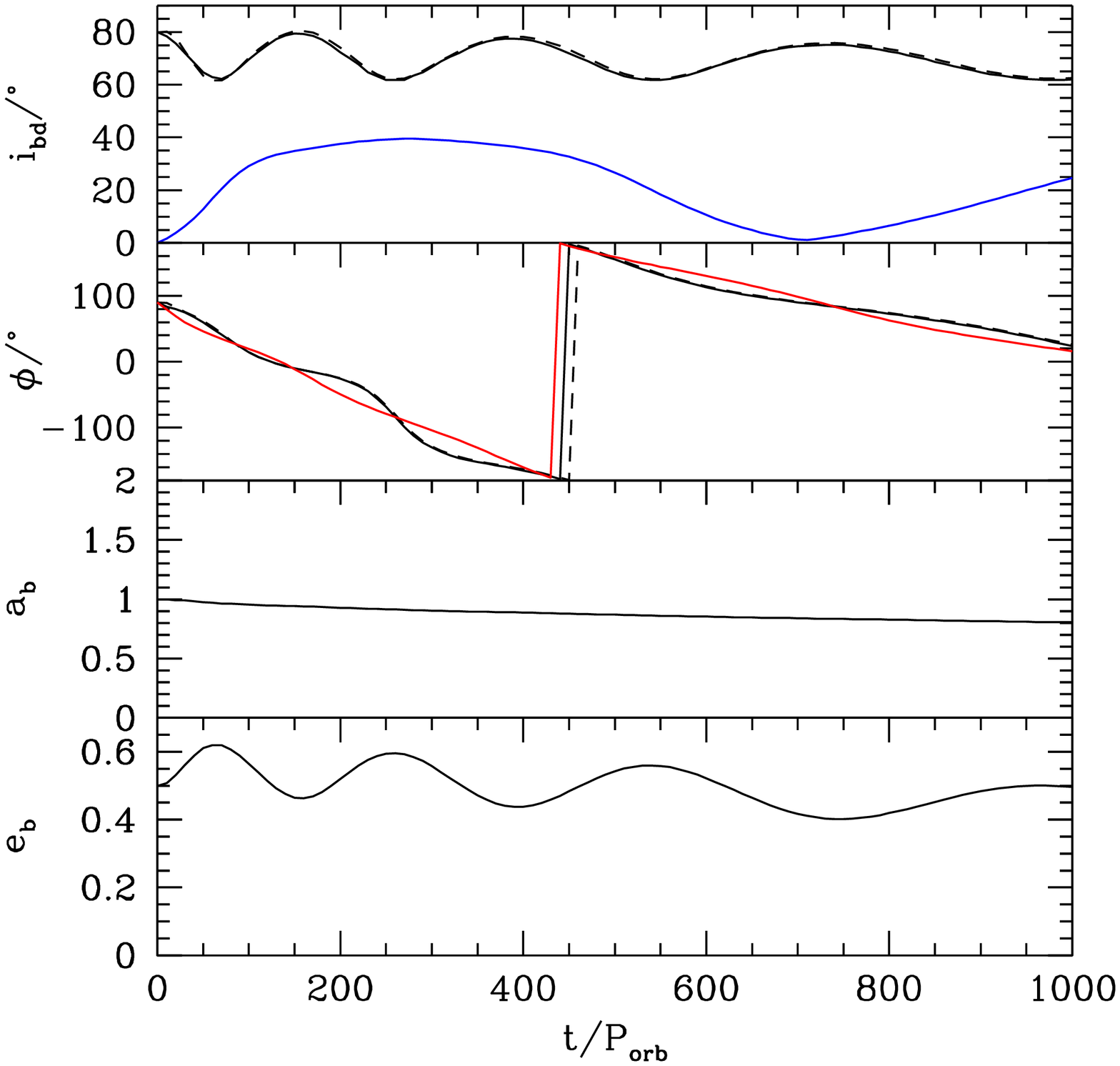}
\caption{The effect of the initial inclination on the evolution of the high mass disc with initial mass $M_{\rm d}=0.05\,M$ and $e_{\rm b}=0.5$ initially. Top left: initial inclination of $20^\circ$  (run8). Top right: initial inclination of $40^\circ$ (run9). Bottom left: initial inclination of $50^\circ$ (run10). Bottom right: initial inclination of $80^\circ$ (run11). }
\label{inc2}
\end{figure*}

Fig.~\ref{inc2} shows the effect of changing the initial inclination of the disc for an initially high disc mass of $0.05\,M$ and an initial binary eccentricity of $0.5$. The simulations that have initial inclination $20^\circ$ (top left, run8), $40^\circ$ (top right, run9) and $50^\circ$ (bottom left, run10) undergo nodal phase circulation of the disc relative to the binary. The disc and the binary are seen to be precessing in opposite directions. However, for initial inclination of $60^\circ$ (bottom right of Fig.~\ref{mass}, run4) and $80^\circ$ (bottom right of Fig.~\ref{inc2}, run11), the disc is librating relative to the binary. The precession angles of the binary and the disc are nearly locked together. Thus, the critical angle between the two types of solution for these parameters is in the range $50-60^\circ$. 
A disc in this librating state is then in a polar-like orbit around the binary.

For the disc with the initial inclination of $80^\circ$, the tilt oscillations are in the opposite direction to the lower inclination discs. In other words, the inclination initially decreases and the eccentricity increases, vice versa for the lower inclination simulations.
The disc is approaching its generalised polar angle  $i_{\rm s}$ from above.

\subsubsection{Binary eccentricity $e=0.8$}

\begin{figure*}
\centering
\includegraphics[width=\columnwidth]{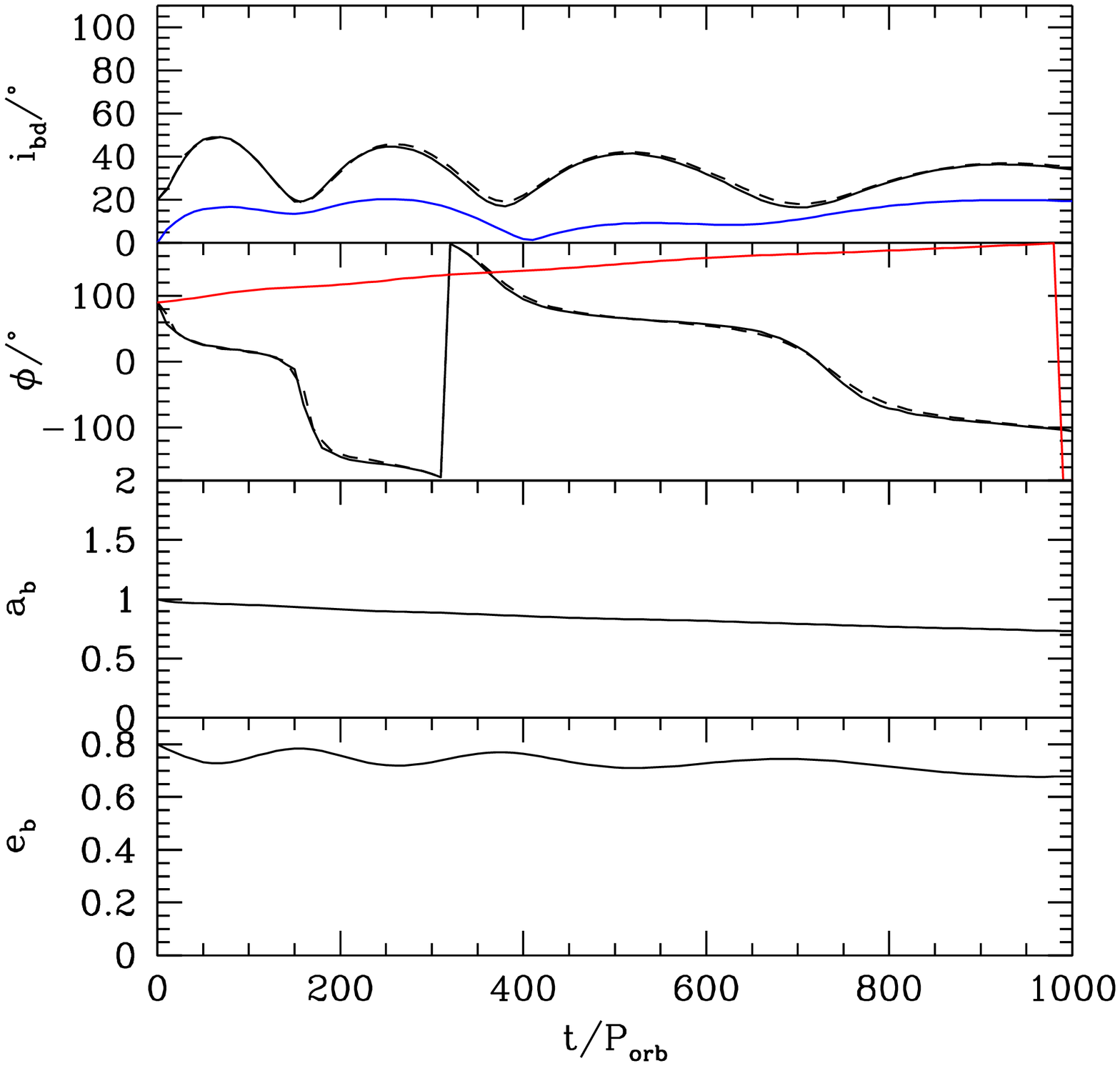}
\includegraphics[width=\columnwidth]{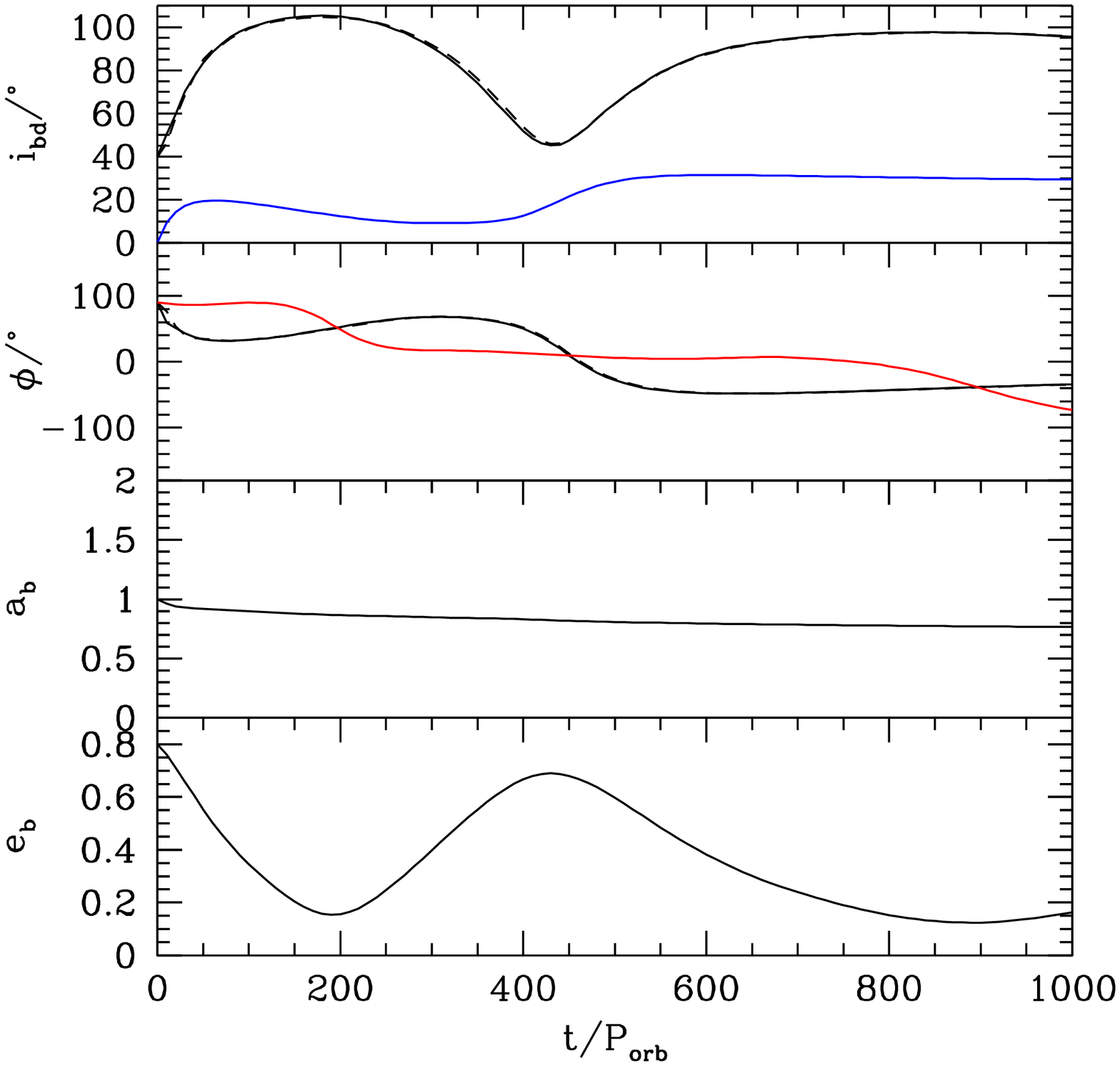}
\includegraphics[width=\columnwidth]{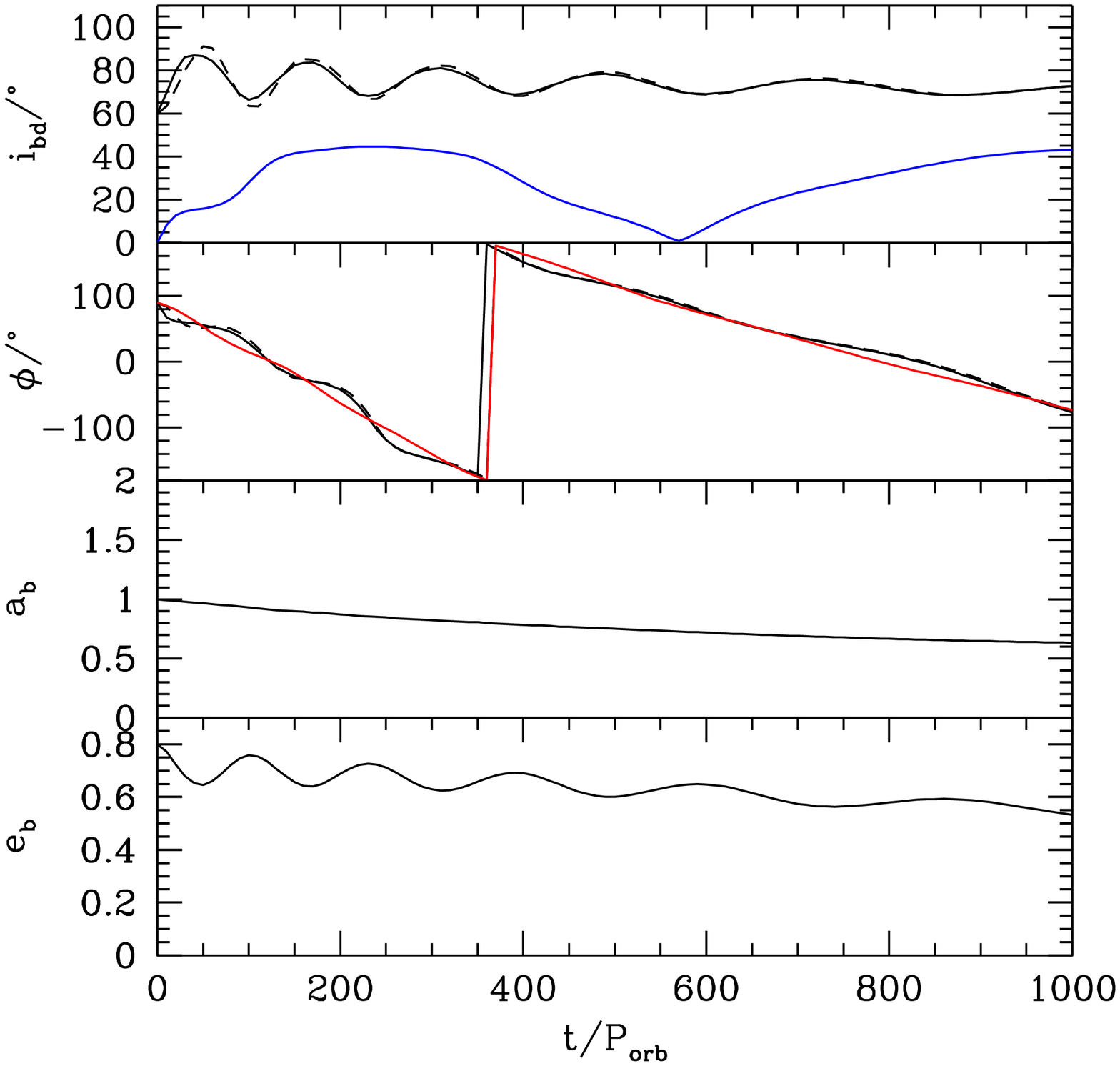}
\includegraphics[width=\columnwidth]{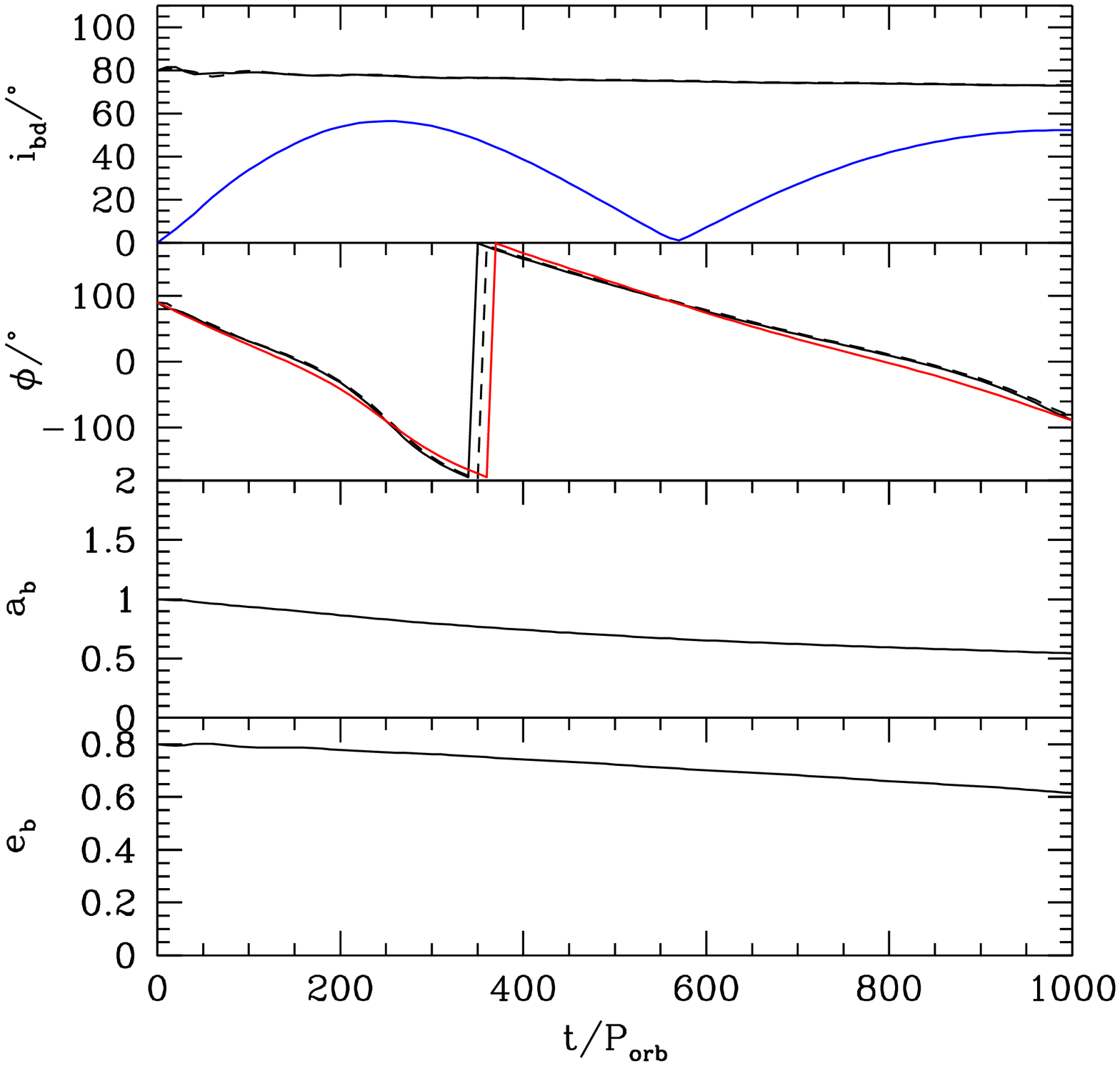}
\caption{The effect of the initial inclination on the evolution of the high mass disc with initial mass $M_{\rm d}=0.05\,M$ and $e_{\rm b}=0.8$. Top left: initial inclination of $20^\circ$  (run12). Top right: initial inclination of $40^\circ$ (run14). Bottom left: initial inclination of $50^\circ$ (run15). Bottom right: initial inclination of $80^\circ$ (run16). }
\label{inc3}
\end{figure*}

Fig.~\ref{inc3} shows the effect of changing the inclination of the disc around a  binary with a higher eccentricity of $e_{\rm b}=0.8$. The disc varies from circulating phase at initial inclination of $20^\circ$ (top left panel, run12) to librating phase for initial inclination $40^\circ$ (top right panel, run14). Although we do not show a figure, we also ran a simulation with an initial inclination of $30^\circ$ and find that it is circulating (see run13 in Table~\ref{tab}). Thus, the critical angle is between $20$ and $30^\circ$.   This angle is higher than the critical angle expected for a test particle of $16^{\circ}$ based on equation 2 of \cite{Doolin2011}. The angular momentum evolution of the simulation that begins at $40^\circ$ (run14) is shown in the short--dashed line in Fig.~\ref{angmom}.

\subsection{Size of the disc}

\begin{figure*}
\centering
\includegraphics[width=\columnwidth]{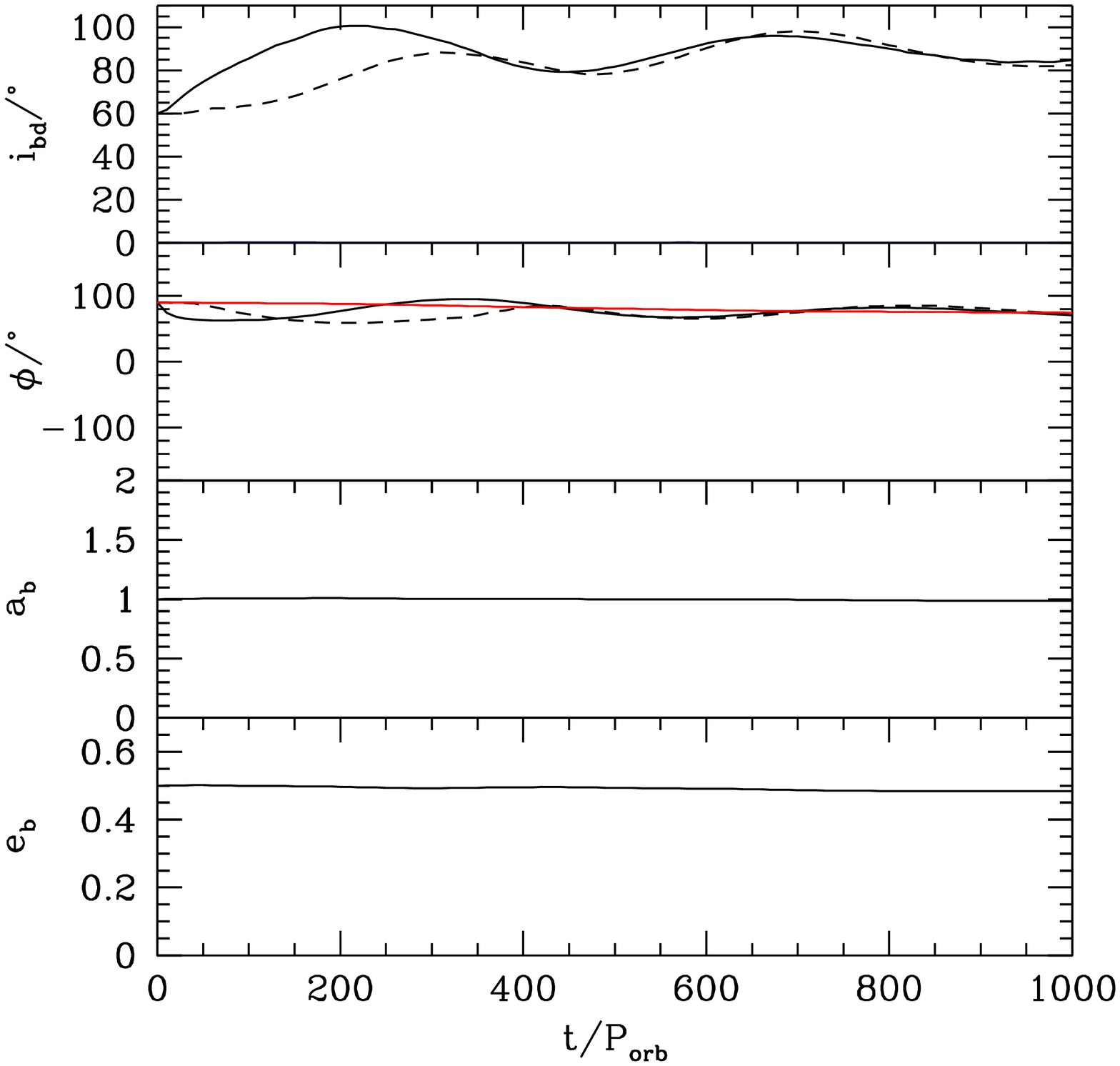}
\includegraphics[width=\columnwidth]{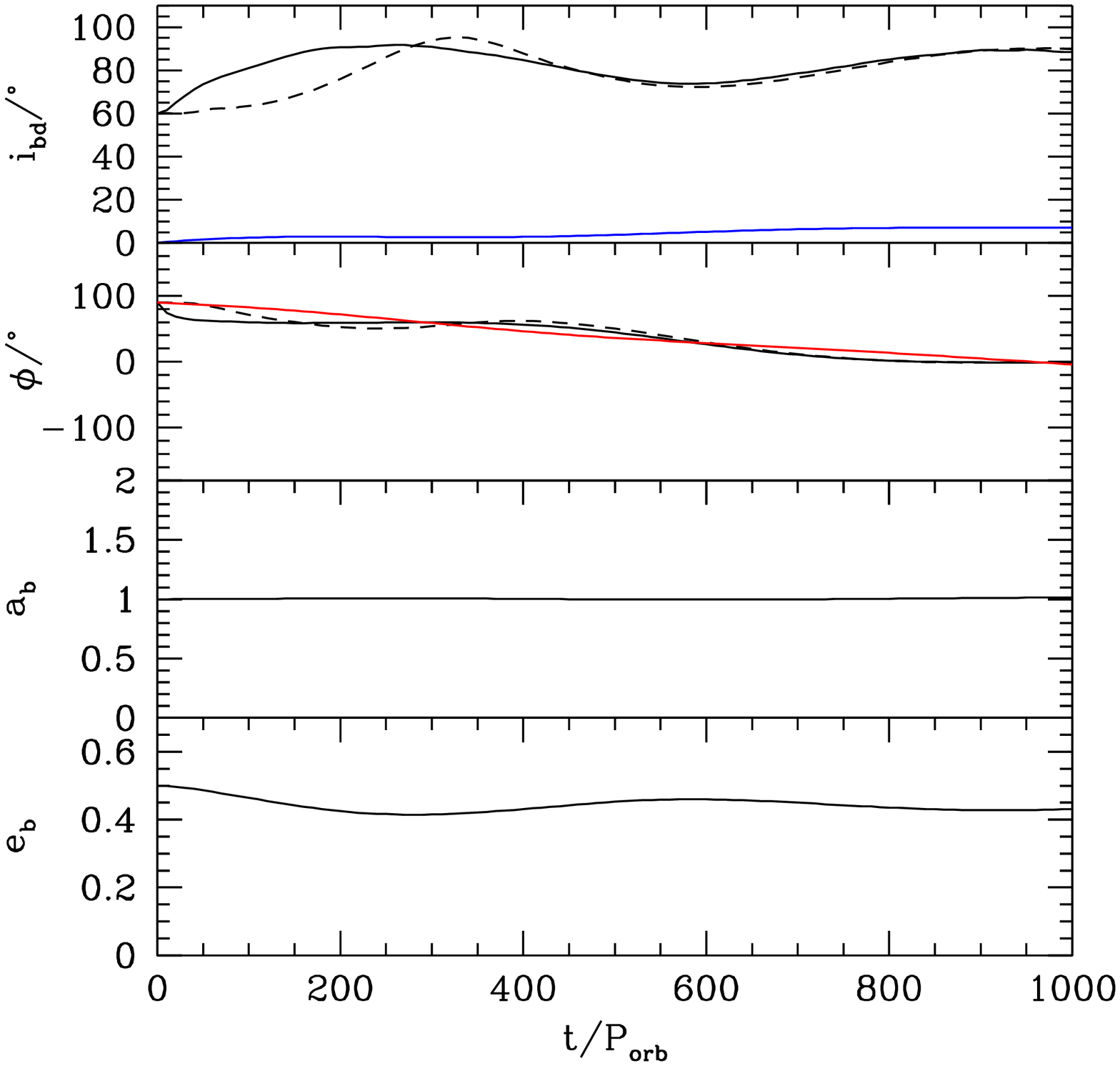}
\includegraphics[width=\columnwidth]{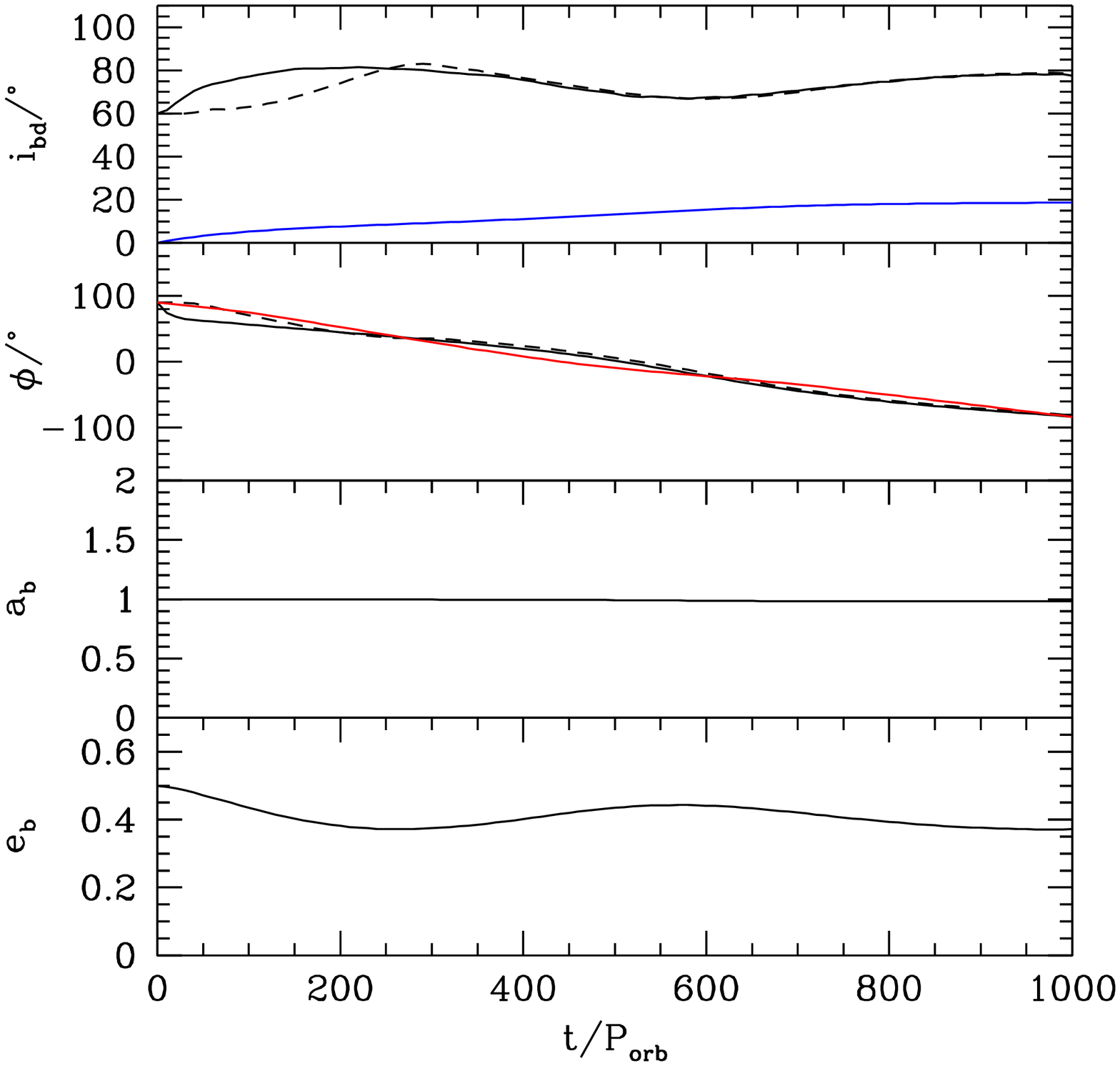}
\includegraphics[width=\columnwidth]{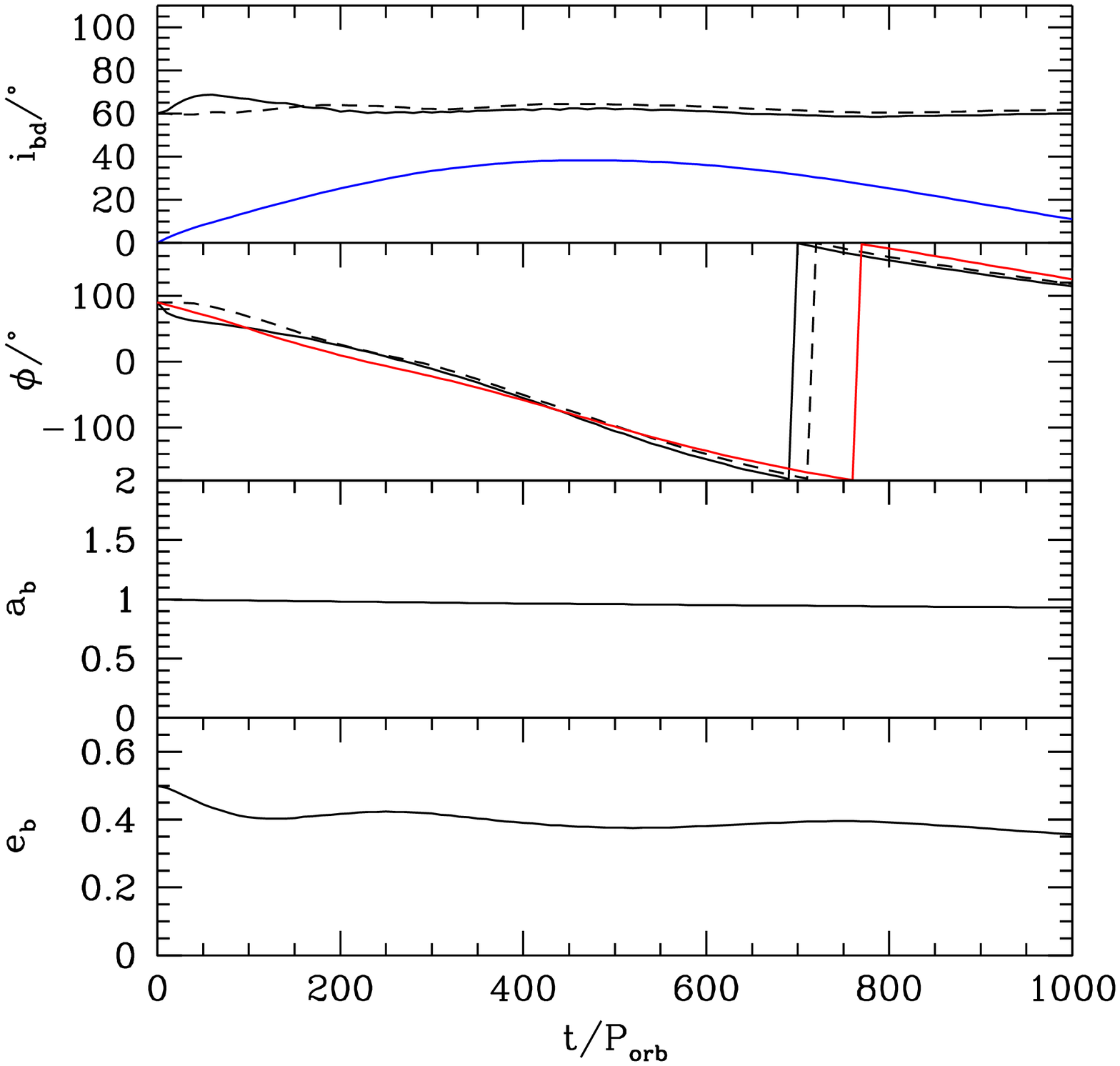}
\caption{Same as Fig.~\ref{mass} except the initial disc outer radius is $10\,a_{\rm b}$. The initial mass of the disc is $0.001\,M$ (top left, run17), $0.01\,M$ (top right, run18), $0.02\,M$ (bottom left, run19) and $0.05\,M$ (bottom right, run20). The solid lines show a radius of $R=3\,a_{\rm b}$ and the dashed lines $R=10\,a_{\rm b}$.  }
\label{inc4}
\end{figure*}

\begin{figure*}
\centering
\includegraphics[width=\columnwidth]{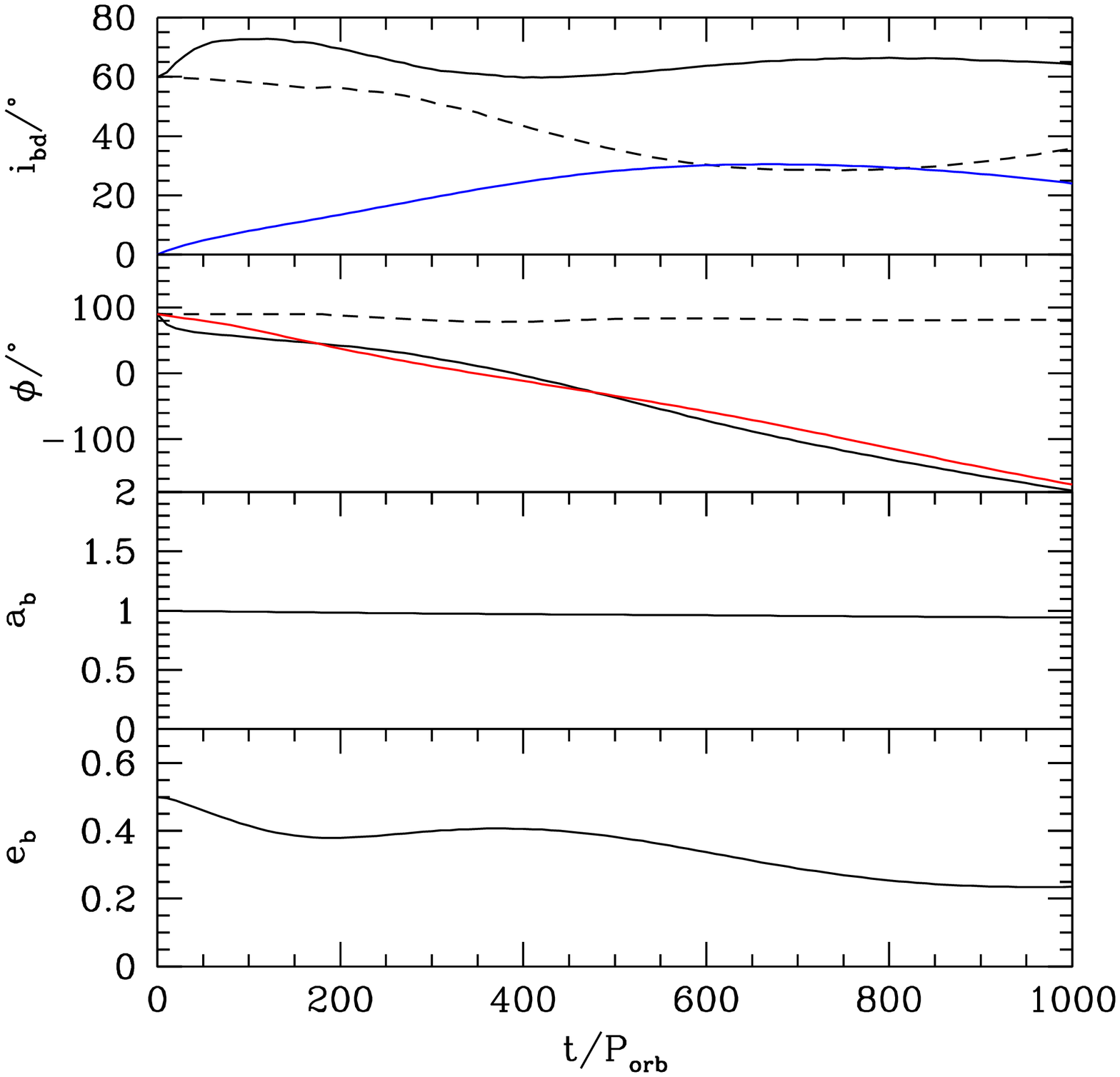}
\includegraphics[width=\columnwidth]{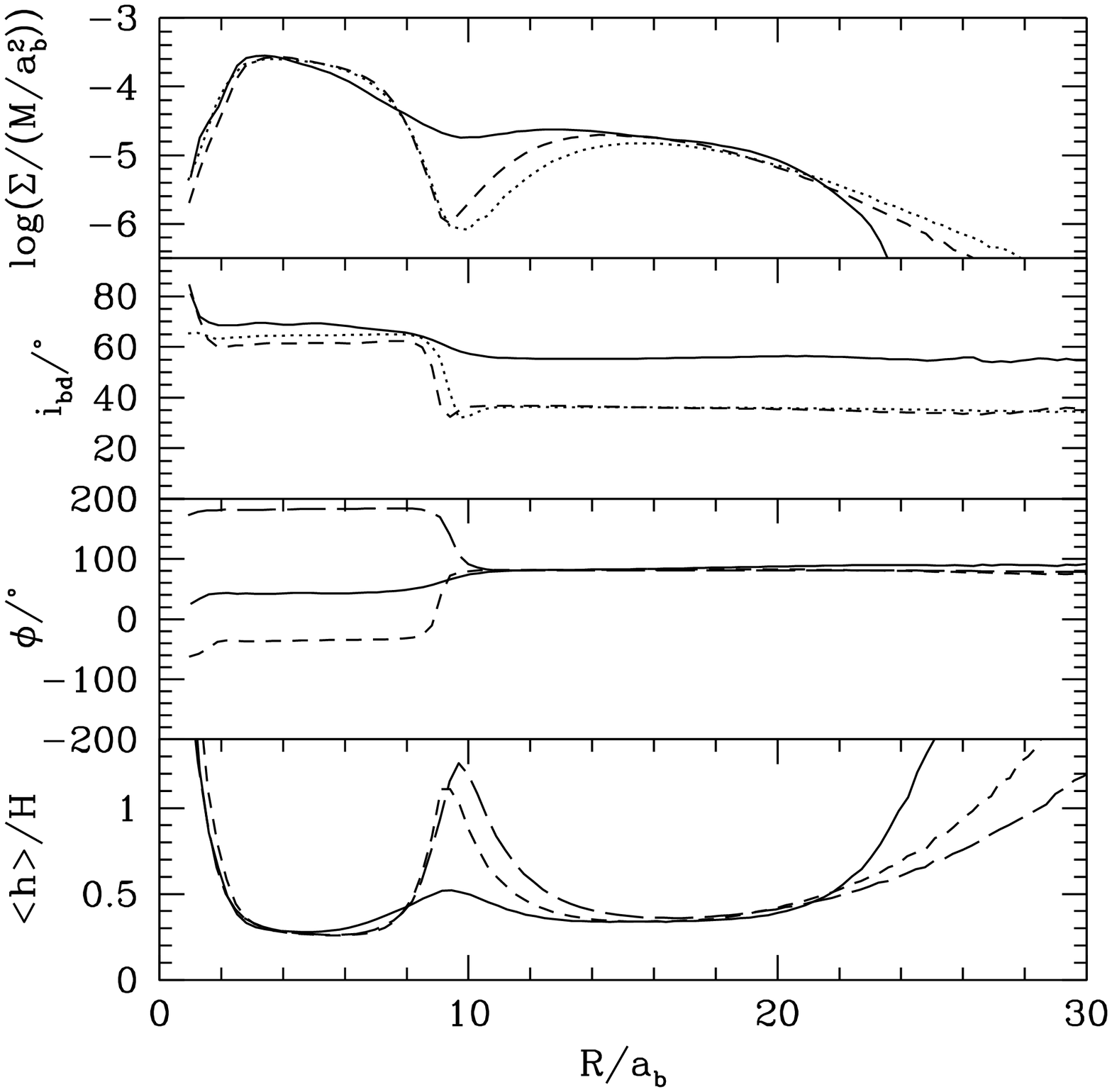}
\caption{Left: Same as the lower right panel of Fig.~\ref{mass} except the initial disc outer radius is $20\,a_{\rm b}$. The initial mass of the disc is $0.05\,M$ (run21). The solid lines show a radius of $R=3\,a_{\rm b}$ and the dashed lines $R=20\,a_{\rm b}$. Right: The surface density, inclination, phase angle,  and smoothing length--to--disc scale height ratio as a function of radius at times $t=200\,P_{\rm orb}$ (solid lines), $t=500\,P_{\rm orb}$ (dashed lines) and $t=1000\,P_{\rm orb}$ (dotted lines).}
\label{inc5}
\end{figure*}

The size of the disc relative to the binary separation may take a wide range of values. Protoplanetary discs are thought to extend to around hundreds of au \citep[e.g.][]{Williams2011}. For a close binary, this may be several hundred binary separations. However, for a wider binary this may be only a few times the binary separation. The simulations we have considered so far in this work have a moderate extent and are relevant to wider binaries. In \cite{Martin2018} we found that extending the outer disc radius, relative to the binary separation led to warped and even broken discs. If the sound crossing timescale over the disc is longer than the precession timescale, then the disc is unable to communicate fast enough to remain as a solid body.  

\subsubsection{Initial disc outer radius $R=10\,a_{\rm b}$}

Fig.~\ref{inc4} shows the effect of increasing the initial size of the disc to $10\, a_{\rm b}$ compared to $5\, a_{\rm b}$ that
we previously described. The figure shows the same four disc masses as shown in Fig.~\ref{mass}.  The qualitative behaviour of the disc has not changed by increasing the initial disc radius. In each case, the disc is in a librating state. The two lines in the inclination and phase angle plots show the disc at a radius of $3\,a_{\rm b}$ (solid lines) and $10 \,a_{\rm b}$ (dashed lines). There is a much more noticeable difference between these two radii now. That is, there is more warping in the larger disc. The warping is larger for the smaller disc mass because the tilt oscillations are larger.   For high mass broader disc, the generalised polar (stationary) inclination is $i_{\rm s}\approx 60^\circ$ that is slightly lower than for the narrower disc. Thus, the disc begins very close to its stationary angle $i_{\rm s}$ and so there is little inclination evolution. 
For the largest disc mass considered (run20), the evolution of the ratio of the disc angular momentum to the binary angular momentum is shown in the long--dashed line in Fig.~\ref{angmom}.  The lower left panel of Fig.~\ref{splash} shows the disc at a time of $t=1000\,P_{\rm orb}$.

\subsubsection{Initial disc outer radius $R=20\,a_{\rm b}$}

The left hand panel of Fig.~\ref{inc5} shows the high disc initial mass case  of $M_{\rm d}=0.05\,M$ with an even larger initial disc outer radius of $20\,a_{\rm b}$ (run21). The two lines in the inclination and phase angle plots show the disc conditions at a radius of $3\,a_{\rm b}$ (solid lines) and $20 \,a_{\rm b}$ (dashed lines). There is significant difference in properties between the two parts of the disc. Hence in the right hand panel we show the surface density, inclination and phase angle as a function of radius at three different times. There is a clear break in the disc at a radius of about $10\,a_{\rm b}$. Circumbinary discs simulations around circular binaries have previously shown this behaviour \citep{Nixonetal2012b,NK2012}. The inner and the outer parts of
the disc precess independently and show tilt oscillations on different
timescales. The inner part of the broken disc at least can still achieve polar alignment.  For this simulation (run21), the evolution of the ratio of the disc angular momentum to the binary angular momentum is shown in the dot--dashed line in Fig.~\ref{angmom}.  

 The lower right panel of Fig.~\ref{splash} shows the broken disc at a time of $t=1000\,P_{\rm orb}$. The inner part of the disc is in a generalised polar aligned state while the outer part remains misaligned. The lower panel on the right hand side of Fig.~\ref{inc5} shows the smoothing length as a function of radius in the breaking disc. At the break, the smoothing length increases because of the small amount of material in the low density gap (see Fig.~\ref{splash}). Disc breaking, as we find, can only be seen in discs with sufficiently high resolution \citep{Nealon2015}.

\section{Generalised polar alignment of a  ring with mass}
\label{sec:polarana}

The secular dynamics
of a circumbinary particle are identical to those of a narrow circumbinary ring.
In order to understand the stable polar alignment of a disc with significant mass, in this section we consider a three body problem for a circumbinary particle that takes into account the gravitational effects of the masses of all three bodies.  We first determine the inclination at the centre of the librating region $i_{\rm s}$, where the ring
(particle) nodal phase is stationary with respect to the binary nodal phase.  We then determine the conditions required for a circumbinary ring to evolve into a stationary (polar)
configuration. The ring model provides insight into the effects gravitational interactions by the orbiting ring. But it does not include possible effects due to the radial extension of a disc or the advection of disc mass and angular momentum on to the binary.

\subsection{Evolution equations}

\cite{Farago2010} developed a secular theory for the motion of a circumbinary
particle of nonzero mass. The principal approximation is that the binary
potential is calculated in the quadrupole approximation. They
utilize a Cartesian coordinate system that is defined relative to the binary orbit. The orbit changes in time due to gravitational interactions with the particle.
The $x$-direction is along the instantaneous
eccentricity vector of the binary, the $z$-direction is along the 
instantaneous binary angular momentum, and the $y$-direction is orthogonal
to the $x$ and $z$ directions. The origin lies at the instantaneous center of mass of the binary. The equations of motion of the particle
are expressed in terms of a unit vector that lies along the direction
of the ring's (particle's) angular momentum that we denote by tilt vector 
$\bm{\ell}=(\ell_{x}, \ell_{y}, \ell_{z})$ in this coordinate system.

As shown by \cite{Farago2010}, the
circumbinary ring semi-major axis, the eccentricity (that we assume to be zero),
and its angular momentum, $J_{\rm r}$,
are constants of motion.
For the binary, the semi-major axis $a_{\rm b}$ is a constant of motion, while its eccentricity, angular momentum $J_{\rm b}$, and binary-ring mutual inclination $i$ are not constants of motion. However, the system angular momentum $J$
is a constant of motion. These properties imply that
\begin{equation}
J_{\rm b}^2 + 2 J_{\rm b} J_{\rm r} \cos{i} = J^2 -J_{\rm r}^2,
\label{CJ}
\end{equation}
where the LHS is a constant of motion.
In this equation, since $a_{\rm b}$ is a constant of motion, binary angular momentum $J_{\rm b}$ varies
in time due to variations in binary eccentricity $e_{\rm b}$ 
 as inclination $i$ varies in time.
This equation then determines a relationship between $e_{\rm b}$
and $i$. In the limit that
$J_{\rm b} \gg J_{\rm r}$, Equation (\ref{CJ}) implies that $J_{\rm b}$
and therefore $e_{\rm b}$ are constants of motion, as applies
for a low mass ring. In the opposite
limit of a very massive ring  $J_{\rm b} \ll J_{\rm r}$,
we have that $J_{\rm b} \cos{i}$
is a constant of motion. This condition holds
because the $z$ component 
of the binary angular momentum is conserved due to the static potential imposed by the massive stationary ring. The constant of motion in this case
plays a key role in the study of Kozai-Lidov oscillations \citep{Kozai1962,Lidov1962}.

The equations of motion track the variations in time of the tilt vector $\bm{\ell}$ and
the binary eccentricity $e_{\rm b}$.
We apply the secular evolution equations 3.15 - 3.18 of  \cite{Farago2010}. We make some changes in variables. 
We also make use of the ratio of the ring-to-binary angular momentum
\begin{equation}
    j = \frac{J_{\rm r}}{J_{\rm b}}. \label{j}
\end{equation}
The angular momentum ratio $j$ generally varies in time because $J_{\rm b}$ varies in time, while $J_{\rm r}$
does not change in time.
 We write the evolution equations as
\begin{align}
\frac{d \ell_x}{d \tau} &= (1- e_{\rm b}^2) \ell_y \ell_z + \gamma_{\rm r} \sqrt{1- e_{\rm b}^2} \,\ell_y (2- 5 \ell_x^2), \label{lx}\\
\frac{d \ell_y}{d \tau} &=- (1+ 4 e_{\rm b}^2) \ell_x \ell_z \nonumber \\  
  & \qquad -\frac{ \gamma_{\rm r} \ell_x }{\sqrt{1-e_{\rm b}^2}} \left( (1-e_{\rm b}^2)  (2- 5 \ell_x^2) + 5 e_{\rm b}^2 \ell_z^2 \right), \label{ly}\\
\frac{d \ell_z}{d \tau} &=  5 e_{\rm b}^2  \ell_x \ell_y +  \frac{5 \gamma_{\rm r}  e_{\rm b}^2}{\sqrt{1-e_{\rm b}^2}}  \ell_x \ell_y  \ell_z, \label{lz}\\
\frac{d e_{\rm b}}{d \tau} &=   5 \gamma_{\rm r} e_{\rm b} \sqrt{1-e_{\rm b}^2}  \ell_x \ell_y, \label{eb}
\end{align}
where we apply a scaled time equal to $\tau= \alpha' t$ for time $t$ in taking the time derivatives above. Quantity $\alpha'$ is constant in time and is defined by equation 3.9 of \cite{Farago2010}.
For our purposes of determining closed orbits. we do not care about the actual time $t$ and therefore do not need to know the value of $\alpha'$.
So we use $\tau$ as our time coordinate.
Quantity $\gamma_{\rm r}$ is proportional to the ring angular momentum and
is a constant of motion 
\begin{equation}
\gamma_{\rm r} = \sqrt{1-e_{\rm b}^2} \, j.
\label{gr1}
\end{equation}
For the purposes of numerically integrating these equations,  it is convenient
to set $\gamma_{\rm r} $ as
\begin{equation}
\gamma_{\rm r} = \sqrt{1-e_{\rm b0}^2} \, j_0,
\label{gr2}
\end{equation}
where $e_{\rm b0}$ and $j_0$
are the initial eccentricity and ring-to-binary angular momentum ratio, respectively.

\subsection{Stationary inclination}
\label{sec:stationary}

We are interested in determining the conditions for $\bm{\ell}$ to be stationary in the $\ell_y=0$ plane. We then require that
\begin{eqnarray}
\frac{d \bm{\ell}}{d \tau}&=& 0, 
\label{statcond1}\\
\frac{d e_{\rm b}}{d \tau} &=& 0 
\label{statcond2}
\end{eqnarray}
in Equations (\ref{lx}) - (\ref{eb}). 
For the test particle case,
we know that this occurs  when the particle orbit lies perpendicular
to the binary orbital plane so that $\bm{\ell}=(1, 0, 0)$.
(It also occurs for $\bm{\ell}=(-1, 0, 0)$ corresponding
to an anti-alignment of particle angular momentum with binary eccentricity.
But we omit discussion of that orientation.)
In the present case, we take into account the nonzero ring mass.
The stationary condition in the $\ell_y=0$ plane is given by
\begin{eqnarray}
   \ell_{x} &=&  \sqrt{1-\ell_{\rm z}^2},  \label{farx}\\
   \ell_{y} &=& 0, \\
 \ell_{z} &=&  \frac{ -(1+ 4e_{\rm b}^2)+ \sqrt{ \left(1+ 4e_{\rm b}^2 \right)^2+60 (1-e_{\rm b}^2) j^2} }{10 j},
    \label{farz}
\end{eqnarray}
as is consistent with Appendix A.4 of \cite{Farago2010}  (see Appendix A).
In this stationary state, the binary eccentricity $e_{\rm b}=e_{\rm b0}$ and  the ring-to-binary angular momentum ratio $j=j_0$ are constant in time. 
From Equation (\ref{farz}) we can obtain the stationary  tilt angle of the ring relative to the binary 
using the fact that 
\begin{equation}
    \cos{i_{\rm s}} =  \ell_{z}.
    \label{isa}
\end{equation}

For  small ring angular momentum, $j\ll1$, we have that
\begin{equation}
    \cos{i_{\rm s}} \simeq \frac{3 j \, (1-e_{\rm b}^2)}{1+4 e_{\rm b}^2}.
\end{equation}
A zero mass stationary ring is then perpendicular to the binary orbital plane, as expected. For arbitrary ring mass, in the limit of high eccentricity close to unity, we have that
\begin{equation}
    \cos{i_{\rm s}} \simeq \frac{6 j\, (1-e_{\rm b})}{5}.
\end{equation}
The stationary tilt angle $i_{\rm s}$ then increases with binary eccentricity.
The stationary angle is achieved at a near perpendicular orientation for sufficiently
large binary eccentricity. 
In the limit of large ring angular momentum $j \gg1$, the stationary inclination is 
\begin{equation}
\cos{i_{\rm s}} \simeq \sqrt{\frac{3}{5}\left(1-e_{\rm b}^2\right)}. 
\label{largej}
\end{equation}
Note that in the case of circular binary orbit, the stationary angle is the critical angle for Kozai--Lidov oscillations of $39.2^\circ$ \citep{Kozai1962,Lidov1962}.

\begin{figure*}
\centering
\includegraphics[width=\columnwidth]{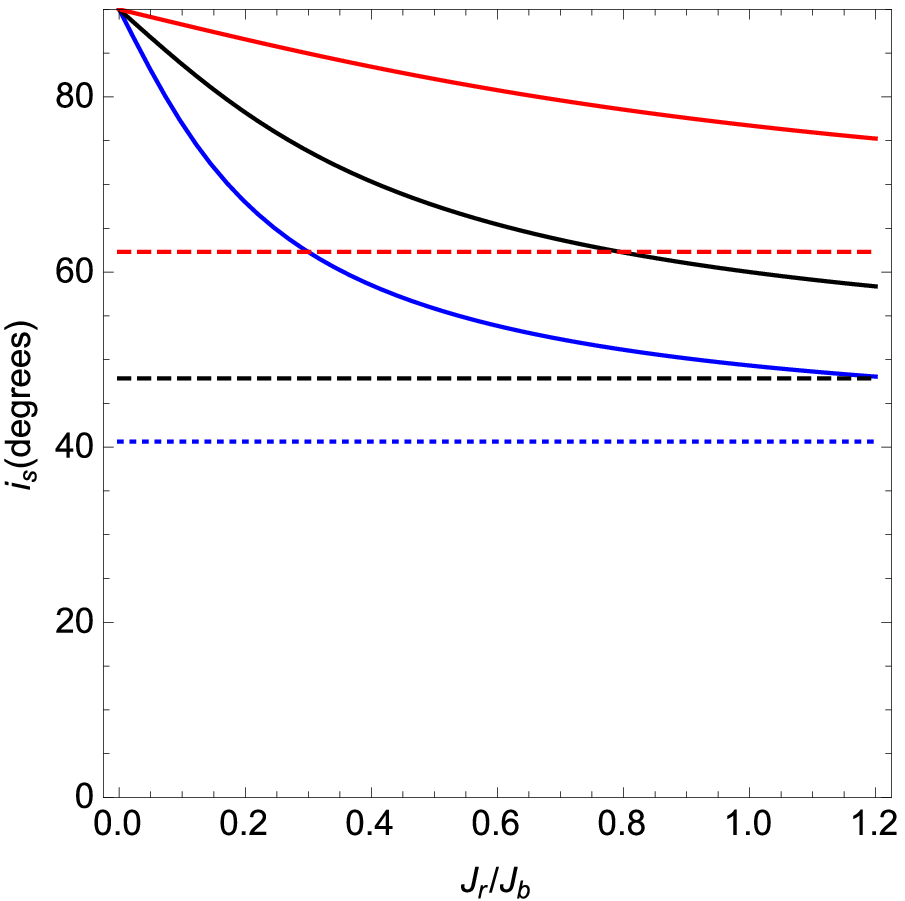}
\includegraphics[width=\columnwidth]{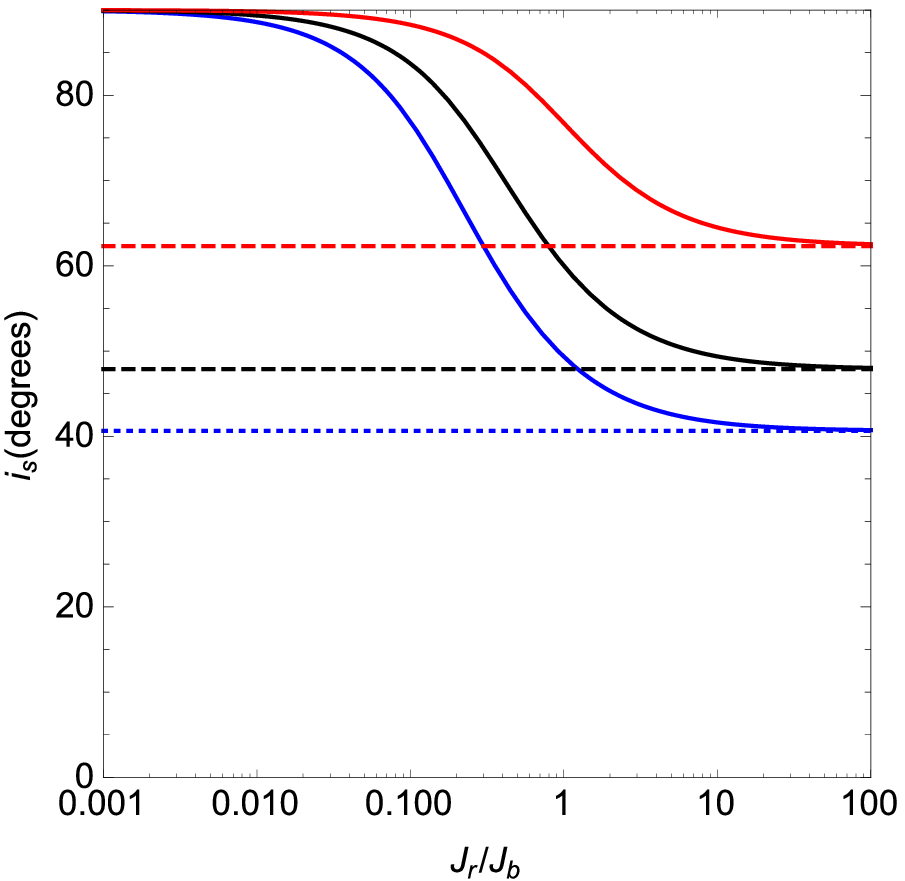}
\caption{The solid lines show the stationary inclination, $i_{\rm s}$, as a function of the ratio of the ring angular momentum to the binary angular momentum for binary eccentricity $e_{\rm b}=0.2$ (blue), $e_{\rm b}=0.5$ (black) and $e_{\rm b}=0.8$ (red) found with Equations~(\ref{farz}) and~(\ref{isa}). The dashed lines show the stationary inclination in the limit that this angular momentum ratio goes to infinity, given by Equation~(\ref{largej}).  The left and right hand panels are the same except the right panel is on a log scale to show the convergence to the dashed lines at large $J_{\rm r}$.}  
\label{fixedpoint}
\end{figure*}

Fig.~\ref{fixedpoint} shows the stationary inclination as a function of the ratio of the ring angular momentum to the binary angular momentum for three different binary eccentricities (using Equations~(\ref{farz}) and~(\ref{isa})). The dashed lines show the corresponding limit of large particle angular momentum given in Equation~(\ref{largej}).   With increasing ring angular momentum and all other parameters fixed, the stationary tilt angle {\it decreases monotonically} to the value in the corresponding dashed line given by Equation~(\ref{largej}) at large $J_{\rm r}/J_{\rm b}$. 
\cite{Zanazzi2018} also found that the stationary tilt (fixed point) is less than $90^\circ$ for a circumbinary particle (ring) with nonzero angular momentum. They obtained numerical results for this problem with different conditions for stationary solutions than our conditions given by Equations~(\ref{statcond1}) and~(\ref{statcond2}). Consequently, our analytic solution (Equations~(\ref{farz}) and~(\ref{isa})) does not agree with their results plotted in their figure 7. We compare our analytic stationary inclination to the numerical hydrodynamical disc simulations in Section~\ref{comp1}.

 If the binary mass ratio decreases (keeping everything else fixed), the angular momentum of the ring  compared to the binary is larger, and therefore $j$ increases. According to Equations (\ref{farz}) and (\ref{isa}), this leads to a lower stationary inclination as seen in Fig.~\ref{fixedpoint}. As noted in \cite{Martin2018}, the libration period also increases with the decreasing binary mass ratio. Thus, the timescale to reach the generalised polar state is also affected by the binary mass ratio.

\subsection{Conditions for polar evolution}
\label{sec:citinc}

We consider here the conditions required for a ring with nonzero mass to evolve towards a stationary
noncoplanar (polar) orientation.
For such evolution to occur, the ring needs to be in a state where its angular momentum direction $\bm{l}$ undergoes libration oscillations
about the stationary  direction
 described in Section \ref{sec:stationary}.

\begin{figure}
\includegraphics[width=\columnwidth]{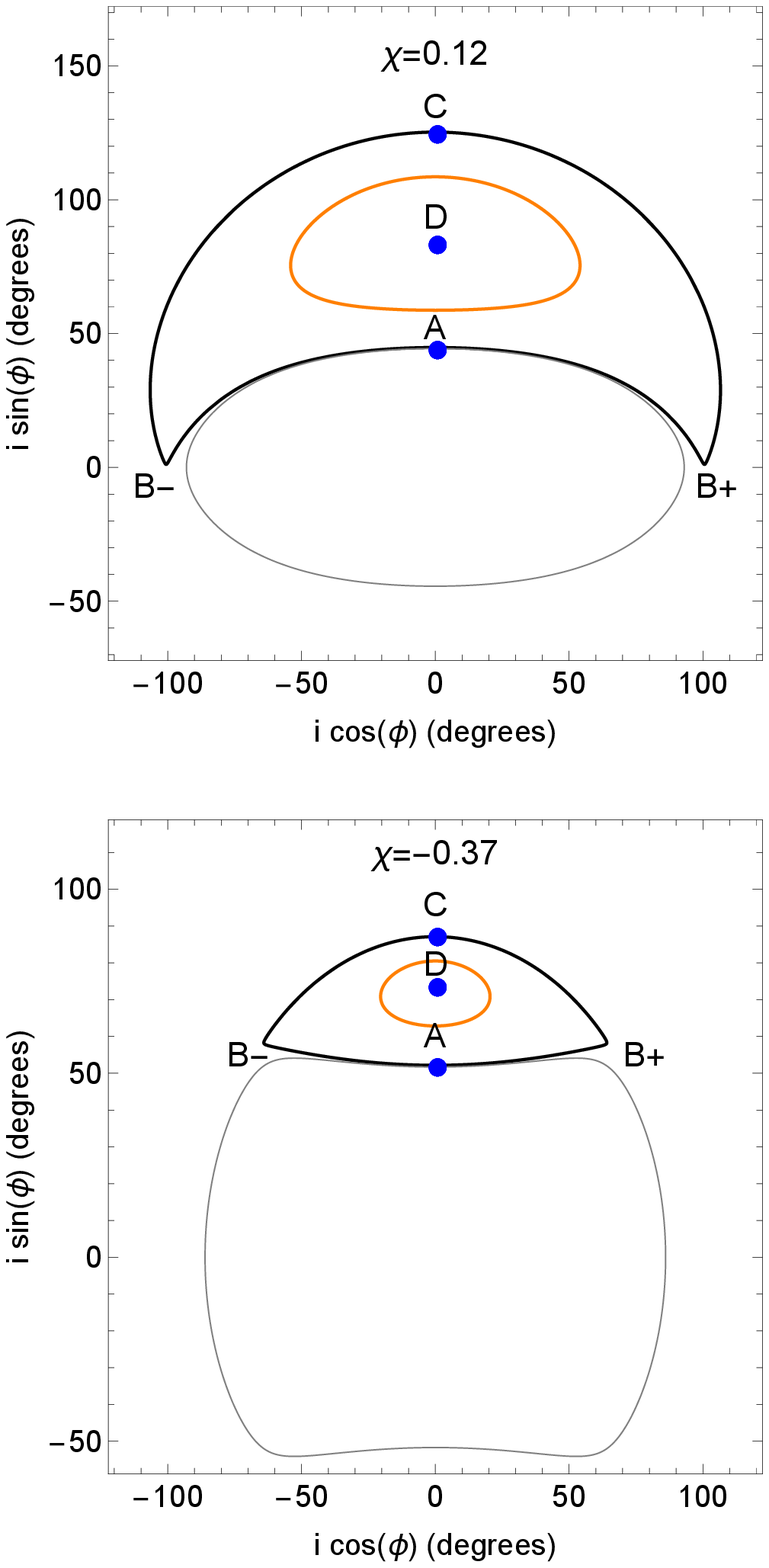}
\caption{ The $i \cos{\phi} - i \sin{\phi}$ plane for orbits with varying initial inclinations and all other parameters held fixed. The black curves are the orbits for the minimum angle
that produces a librating orbit. The grey curves are for orbits whose inclination is slightly lower and are circulating. Point $A$ on the black line denotes the point that has the minimum angle
for a librating orbit. Points $B _{\pm}$ on the black line denotes the location where there is a cusp and the velocity $d\bm{l}/d \tau$ vanishes.
Point $C$ on the black line  is the location of the maximum tilt angle of a librating orbit. Point $D$  marks the location of the stationary (nonlibrating) orbit that
corresponds to a fixed polar orbit relative to the binary. The orange line plots an intermediate librating orbit.  The upper plot has a positive value of $\chi$, while the lower plot has a negative value (see Equation (\ref{lambda})) which indicates a change in orbit behaviour.
 }
\label{fig:CritOrb}
\end{figure}

We determine the minimum inclination
required for a librating orbit, given the binary eccentricity and a measure of the ring-to-binary angular momentum $j$. Since this ratio varies in time as $J_{\rm b}$ varies, we select a value of $j=j_0$ where the line of ascending notes is equal to $\phi=90^\circ$ and the inclination is smaller than the stationary value $i_{\rm s}$. The latter condition is applied because a librating 
orbit of $\bm{l}$ forms a closed loop that is double valued in $\phi$
corresponding to two different values of inclination $i$ (see points $A$ and $C$ in Figure~\ref{fig:CritOrb}). We select
the $j$ value at the smaller value of $i$ ($i < i_{\rm s}$) for the reference quantity $j_0$.  Similarly binary eccentricity $e_{\rm b}$ varies
in time and we apply the value of the reference binary eccentricity
$e_{\rm b0}$ that is the value of eccentricity at this same phase $\phi=90^\circ$ and inclination. 

For a given initial value of binary eccentricity $e_{\rm b0}$, angular momentum ratio $j_0$, and assumed initial binary-ring inclination $i_0 < i_{\rm s}$,
we integrate the evolution Equations (\ref{lx}) - (\ref{eb}) together with Equation (\ref{gr2}). The initial conditions are given by
\begin{equation}
(\ell_x, \ell_y, \ell_z, e_{\rm b})= (\sin{i_0}, 0, \cos{i_0}, e_{\rm b0}).
\label{icond}
\end{equation}
For a fixed set of values of $e_{b0}$ and $j_0$, we determine
the minimum value of $i_0$ for which the orbit of $\bm{\ell}$ is librating, rather than circulating. This is done using a bisection method.
We sometimes refer to that librating orbit as the critical orbit.

 Figure~\ref{fig:CritOrb} plots two critical librating orbits as heavy black lines in a phase portrait of $i \cos{\phi}$ versus $i \sin{\phi}$.
 The distance from the origin to a point on the plot is the mutual inclination $i$, while the angle from horizontal
 to the line from the origin to a point on the plot is equal to the longitude of ascending node $\phi$.
Both plots are for a system with $e_{\rm b0}=0.5$. The upper plot has $j_0=0.1$, while the lower plot has $j_0=0.3$.
The gray lines plot the circulating orbits that result from slightly smaller values of $i_0$ than for the critical orbit.
Both the minimum and maximum values of $i$ along an orbit always occur where $\phi=90^\circ$ corresponding to points $A$ and $C$ in Figure~\ref{fig:CritOrb}. The minimum inclination
along a librating orbit thus occurs at the initial time   when
$i=i_0$ (as described above in Equation (\ref{icond})).

\begin{figure}
\includegraphics[width=\columnwidth]{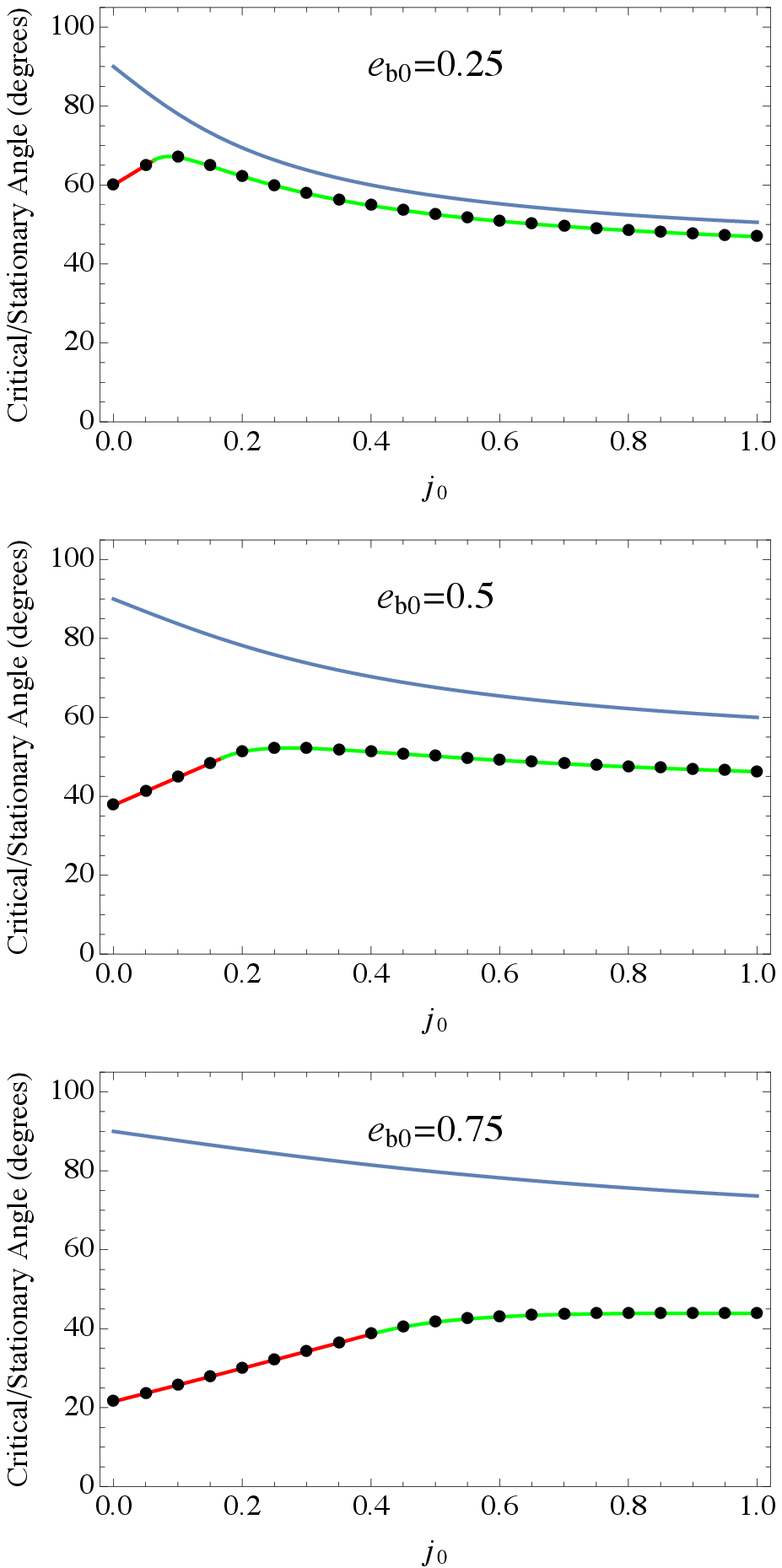}
\caption{Plots of minimum and stationary tilt angles as a function of the ratio of ring-to-binary angular momentum $J_{\rm r}/ J_{\rm b} $ at phase $\phi=90^\circ$ (denoted by $j_0$) and binary eccentricity at this phase $e_{\rm b}=e_{\rm b0}$. 
The dotted lines are the numerically determined  minimum tilt angles for libration which leads to generalised polar alignment. 
The  solid blue lines are the tilts for the
generalised polar alignment (stationary points given by Equations (\ref{farz}) and (\ref{isa}))). 
The solid red lines are an analytic determination of the minimum tilt angle based on Equation (\ref{minlib}) over a range of $J_{\rm r}/J_{\rm b}$ where $\chi >0$
in Equation (\ref{lambda}). The solid green lines are an analytic determination of the minimum tilt angle based on Equation (\ref{minlib2}) over a range of $J_{\rm r}/J_{\rm b}$ where $\chi < 0$. The red and green lines meet at $\chi=0$.}
\label{fig:CritINumerical}
\end{figure}

Figure~\ref{fig:CritINumerical}  plots as a dotted line the numerically determined minimum 
tilt angles for critical librating orbits as a function of $j_0$ for three different values
of binary eccentricity $e_{\rm b0}$. In addition,
the solid blue line plots the stationary angle. Notice that the stationary angles lie above the minimum tilt angles,
as expected (point $D$ in Figure~\ref{fig:CritOrb} lies above point $A$). 
The minimum tilt angle increases  with $j_0$
for small values $j_0$. It flattens and  then decreases for larger values for $j_0$.
If a ring tilt lies above the minimum value given in Figure \ref{fig:CritINumerical},
it does not immediately follow that the ring is in a librating state. The full condition also involves the phase $\phi$ as we see below.
The condition on tilt is necessary for libration,  but not sufficient.

\subsubsection{Lower $j$ / higher $e_{\rm b}$ branch}
\label{lowj}

For sufficiently small values of $j$ or large values of $e_{\rm b}$, it is possible to determine the libration conditions analytically.
As seen in the upper panel of Figure~\ref{fig:CritOrb}  for $j$ small, the critical orbit of $\bm{\ell}$ that separates
libration from circulation has a cusp at $\phi=0^\circ$ and $180^\circ$ that corresponds to $\ell_x=0$ (see points $B_\pm$). The cusp 
involves $d \bm{\ell}/d \tau=0$ on the
$\ell_x=0$ plane  (see also Appendix A3 of \cite{Farago2010}).  Strictly speaking this is a stationary point. But,
this stationary point is unstable, unlike the stationary point in the $\ell_y=0$ plane corresponding to polar configuration discussed in Section \ref{sec:stationary}.
Since it is unstable, orbits that
lie extremely close to it will diverge away from it, either as a librating or circulating orbit.
The librating orbit that comes infinitesimally close to a stationary point with the same binary eccentricity is the  critical librating orbit.

From Equation (\ref{lx}), we have that this $\ell_x=0$  stationary point satisfies
\begin{equation}
  \ell_{z {\rm s}} = - \frac{2 \gamma_{\rm r}}{\sqrt{1- e_{\rm bs}^2}}
  \label{lzs}
\end{equation}
at this stationary point $s$.
From Equations (\ref{CJ}) and (\ref{gr1}), we have that total angular momentum conservation implies
\begin{equation}
 \gamma^2 = \gamma_{\rm r}^2 + 1-e_{\rm b}^2 + 2 \gamma_{\rm r} \sqrt{1-e_{\rm b}^2} \ell_z ,
\label{g1}
\end{equation}
where 
\begin{equation}
 \gamma = \frac{\sqrt{1-e_{\rm b}^2} \, J_{\rm r}}{J_{\rm b}}.
 \label{g2}
\end{equation}
Both $\gamma_{\rm r}$ and $\gamma$ are constants of motion. The reason is that the magnitude
of the binary angular momentum varies in time due the variations in its eccentricity only. The quantity $\sqrt{1-e_{\rm b}^2}/J_{\rm b}$
is then independent of time. We want to express  $\bm{\ell}$ and $e_{\rm b}$ at this stationary point as a function of these constants of motion. The reason
is that they can then be determined from the values of  $\bm{\ell}$ and $e_{\rm b}$ at any point along the critical librating orbit. In this way, the value of the Hamiltonian
at this stationary point can be determined in terms of these values anywhere along the  critical librating orbit by using Equations (\ref{gr1}) and (\ref{g1}).

By using Equation (\ref{lzs}) and applying Equation (\ref{g1}) to this stationary point,
we have that 
\begin{equation}
 e_{\rm bs} = \sqrt{1- \gamma^2 - 3 \gamma_{\rm r}^2}
 \label{ebs}
 \end{equation}
 and 
 \begin{equation}
  \ell_{z {\rm s}} = -\frac{2 \gamma_{\rm r}}{\sqrt{\gamma^2 + 3 \gamma_{\rm r}^2}}
  \label{lzs1}
 \end{equation}
 which are the same as equations A12 and A15 of \cite{Farago2010}.

We
consider a secular Hamiltonian
based on equation 3.21 of \cite{Farago2010}
\begin{equation}
    H = \ell_z^2+ e_{\rm b}^2(2- \ell_z^2-5 \ell_x^2),
    \label{H}
\end{equation}
where we ignore an overall factor that is independent of $\bm{\ell}$ and $e_{\rm b}$ that is irrelevant to our considerations below.

The value of the Hamiltonian
at this stationary point in the $\ell_x=0$ plane based on Equations (\ref{ebs})   and (\ref{lzs1})  is equal to
\begin{equation}
    H_{\rm s} =  2 (1- \gamma^2-  \gamma_{\rm r}^2).
\end{equation}
By applying Equations (\ref{gr1}) and (\ref{g1}), we then have that at any point on the critical librating orbit that
has a Hamiltonian value infintesimally close to $H_{\rm s}$
\begin{equation}
    H_{\rm cr} =  2 \left(e_{\rm b}^2-2 (1-e_{\rm b}^2) j( j + \cos{i}) \right ),
    \label{Hcr}
\end{equation}
where we use the fact that  $\ell_z=\cos{i}.$
But this equation only holds if $e_{\rm bs}$ is real in Equation (\ref{ebs}),
which again through the application of Equations (\ref{gr1}) and (\ref{g1}),
implies that
\begin{equation}
\chi=e_{\rm b}^2 - 2 (1-e_{\rm b}^2) j (2 j + \cos{i}) > 0.
\label{lambda}
\end{equation}
But $\chi$ is a constant of motion and so this equation also holds at any point on the critical orbit.
For negative values of $\chi$, the stationary point in the $\ell_x=0$ plane does not exist.
The condition $\chi >0$ is satisfied for
sufficiently small
values of $j$ or high values of $e_{\rm b}$ close to unity.

When $\chi >0$ we have that the $d \bm{\ell}/d \tau=0$ on the $\ell_x=0$ plane. This means that
the orbit has a cusp for $\phi=0^\circ$ and $180^\circ$. For $\chi < 0$, there is still a cusp
on the critical librating orbit. However, the cusp does not occur
on the $\ell_x=0$ plane.   Figure~\ref{fig:CritOrb} plots the orbits of two cases with different values of $\chi$.
Both plots are for a system with $e_{\rm b0}=0.5$. The upper plot has $j_0=0.1$, while the lower plot has $j_0=0.3$. The middle panel of Figure~\ref{fig:CritINumerical}
shows the critical angles for these cases.

The upper plot of Figure~\ref{fig:CritOrb}  that has $\chi > 0$ has cusp points at $\phi=0^\circ$ and $180^\circ$ as expected.
The critical librating orbit then covers a large $180^\circ$ range of $\phi$. 
The lower plot with $\chi < 0$ has cusp points   (denoted as points $B_\pm$)
that cover a much smaller range in $\phi$.

Libration requires that $H < H_{\rm cr}$. We then obtain
the condition
\begin{equation}
 \Lambda_1 = -(1-e_{\rm b}^2) (2 j + \cos{i})^2 + 5 e_{\rm b}^2 \sin^2{i} \sin^2{\phi} > 0,
 \label{Lambda}
\end{equation}
where we use $\ell_x= \sin{\phi} \sin{i}$ and $\ell_z= \cos{i}$ in evaluating $H$. 
For a massless ring we have that $j=0$ and we recover equation 51 of \cite{Zanazzi2018}.

 The minimum possible tilt for libration occurs where $\phi=90^\circ$. To find the minimum possible tilt $i$ for libration to occur given values of $e_{\rm b0}$ and $j_0$, we use Equation (\ref{Lambda})
with $\phi=90^\circ$ and $\Lambda_1=0$ to obtain
\begin{equation}
    \cos{i_{\rm min}} = \frac{\sqrt{5} e_{\rm b0} \sqrt{4 e_{\rm b0}^2 -4j_0^2(1-e_{\rm b0}^2)+1}- 2j_0(1-e_{\rm b0}^2)}{1+4e_{\rm b0}^2}.
    \label{minlib}
\end{equation}
For small $j_0$, the above equation can be expanded as a series to linear order in $j_0$
to give
\begin{equation}
     \sin{i_{\rm min}} \simeq \sqrt{\frac{1-e_{\rm b0}^2}{1+4e_{\rm b0}^2}} \left(1+ \frac{2 \sqrt{5} e_{\rm b0} \, j_0 }{\sqrt{1+4e_{\rm b0}^2}} \right).
     \label{mintilt}
\end{equation}
For a massless ring we have that $j_0=0$ and $e_{\rm b}=e_{\rm b0}$ (constant binary eccentricity), and  we recover the minimum tilt angle given in equation  (2) of \cite{Doolin2011}.
The term proportional to $j_0$ shows that the minimum
angle for libration increases with ring angular momentum for small $j_0$.

For high binary eccentricity, $e_{\rm b0} $ close to unity, we have to lowest order in $1-e_{\rm b0}$ that
\begin{equation}
 \sin{i_{\rm min}} \simeq \sqrt{\frac{2}{5}} (1+2 j_0) \sqrt{1-e_{\rm b0}}.
\end{equation}
For large $j_0$ in this equation, libration can occur over a wide range of tilt angles provided that $e_{\rm b0} \ga 1-5/(8 j_0^2).$

\subsubsection{Higher $j$ / lower $e_{\rm b}$ branch}
\label{highj}

The results in Section \ref{lowj} are based on a critical librating orbit that
passes through a stationary point (where $d \bm{\ell}/d \tau=0$) that has the property that
$\ell_x=0$. These results apply for $\chi \ge 0$  in Equation (\ref{lambda}) which is based
on the requirement that the binary eccentricity be a real quantity in Equation (\ref{ebs}).
For larger $j$ or lower  $e_{\rm b}$ values,
where $\chi < 0$,
there does not exist a stationary point with $\ell_x=0$. Instead, as we show below,
the critical librating orbit for such larger $j_0$ cases involves a stationary point with
the property that $e_{\rm bs}=0$. The properties of this stationary point are also discussed
in section A2 of \cite{Farago2010}.  

From Equation (\ref{g1}) with  $e_{\rm bs}=0$,
 we have that
\begin{equation}
\ell_{z {\rm s}}= \frac{\gamma^2 -\gamma_{\rm r}^2 -1}{2 \gamma_{\rm r}}.
\end{equation}
Using equation (\ref{H}) for the Hamiltonian, we have that its value at this stationary point  for the
critical librating orbit is then
\begin{equation}
H_{\rm cr}= \left ( \frac{\gamma^2 -\gamma_{\rm r}^2 -1}{2 \gamma_{\rm r}} \right)^2.
\label{Hcr2}
\end{equation}
We proceed as in Section~\ref{lowj}  through the use of Equations (\ref{gr1}) and (\ref{g1})
in Equation (\ref{Hcr2}) and require that $H < H_{\rm cr }$ for a librating orbit to obtain
the condition that
\begin{equation}
\Lambda_2 = e_{\rm b}^2  + 4 j (1-e_{\rm b}^2) \left( -\cos{i} + j (-2 + 5 \sin^2{i} \sin^2{\phi}) \right) > 0,
\label{Lambda2}
\end{equation}
which is valid for $\chi < 0$ in Equation (\ref{lambda}).
The condition $\chi <0$ is satisfied for
sufficiently large
values of $j$ or small values of $e_{\rm b}$.

To find the minimum possible tilt $i$ for libration to occur given values of $e_{\rm b0}$ and $j_0$, we use Equation (\ref{Lambda2})
with $\phi=90^\circ$ and $\Lambda_2=0$ to obtain
\begin{equation}
    \cos{i_{\rm min}} = \frac{ \sqrt{(1- e_{\rm b0}^2) \left(1+4e_{\rm b0}^2 +60 (1-e_{\rm b0}^2) j_0^2 \right)} -(1 - e_{\rm b0}^2) }{10 (1-e_{\rm b0}^2) j_0}.
    \label{minlib2}
\end{equation}
In the limit of large $j_0$, we have to first order
\begin{equation}
  \cos{i_{\rm min}} \simeq \sqrt{\frac{3}{5}}-\frac{1}{10 j_0},
  \label{ijlarge}
 \end{equation}
 which approaches the critical angle for Kozai-Lidov oscillations as $j_0$ goes to infinity.
 
 \subsubsection{Summary of analytic conditions for polar alignment}
\label{sec:anasum}

Suppose we have a system with
the following parameters at some instant in time:
ring-to-binary angular momentum ratio $j=J_{\rm r}/J_{\rm b}$,
mutual binary-ring inclination $i$,
and longitude of ascending node for the ring $\phi$.
If $\chi >0$ in Equation (\ref{lambda}), then Equation (\ref{Lambda}) determines whether the system undergoes libration.
If $\chi < 0$, then Equation (\ref{Lambda2}) determines whether the system undergoes libration.
Libration in turn can lead to alignment
to a stationary (polar) configuration.

The dotted lines in Figure~\ref{fig:CritINumerical} are the minimum tilt angles for libration $i_{\rm min}$ obtained by numerically integrating the tilt evolution Equations (\ref{lx}) - (\ref{eb}).  (The minimum possible tilt for libration  occurs where angle $\phi=90^\circ$.)
The values of $i_{\rm min}$  given analytically by Equation (\ref{minlib}) are plotted with solid red lines in Figure~\ref{fig:CritINumerical} over the range of $j_0$  values  where $\chi >0$  in Equation (\ref{lambda}). 
 The values of $i_{\rm min}$ given analytically by Equation (\ref{minlib2}) are plotted with solid green lines in Figure~\ref{fig:CritINumerical} over the range of $j_0$  values  where $\chi <0$.
 Notice that the red and green lines pass through the dotted lines, indicating excellent numerical agreement between the two independent methods (numerical and analytic) for determining minimum angles
for both branches.
Notice also that there is a change in behaviour of the minimum tilt angle near the largest value of $j_0$ plotted
in red that occurs where $\chi=0$.  Figure~\ref{fig:CritOrb} shows a major
change in orbital behaviour for the librating orbits with a change in sign of $\chi$.
Figure~\ref{fig:CritINumerical} shows that for $\chi <0$ (beyond the red lines),  the minimum tilt does not increase as rapidly with $j_0$ and  decreases for sufficient large $j_0$, as is also indicated by Equation (\ref{ijlarge}). 

\subsection{Comparison of the analytic criteria to the hydrodynamical simulations}
\label{compsa}

\subsubsection{Stationary inclination}
\label{comp1}

Fig.~\ref{angmom} shows the ratio of the disc angular momentum to the binary angular momentum for some of the hydrodynamical simulations.  The high mass disc simulations ($M_{\rm d}=0.05\,M$) with binary eccentricity $e_{\rm b}=0.5$ have initially $J_{\rm d}/J_{\rm b}=0.42$. In the analytic model in Fig.~\ref{fixedpoint}, this corresponds to a stationary inclination of  $i_{\rm s}=69.7^\circ$. This is close to the inclination that the hydrodynamic discs oscillate about (as shown the bottom right panel of  Fig.~\ref{mass} for initial inclination $60^\circ$ and the bottom right panel of Fig.~\ref{inc2} for initial inclination of $80^\circ$). 

For the larger eccentricity binary, $e_{\rm b}=0.8$, with the same disc parameters (Fig.~\ref{inc3}), the system has angular momentum ratio initially $J_{\rm d}/J_{\rm b}=0.61$ and in the analytic model in  Fig.~\ref{fixedpoint}, this corresponds to a stationary inclination of $i_{\rm s}=80.6^\circ$. This is in good agreement with the simulations shown in the top right and bottom left and right panels of  Fig.~\ref{inc3} that are librating.

The simulation with the larger disc size (outer radius $10\,a$) in the bottom right panel of Fig.~\ref{inc4} (run20), has initially $J_{\rm d}/J_{\rm b}=0.52$. In the analytic model this corresponds to a stationary inclination of $i_{\rm s}=67.1^\circ$. The largest disc size we considered (outer radius $20\,a$) in Fig.~\ref{inc5} (run21), has initially $J_{\rm d}/J_{\rm b}=0.68$ and this corresponds to $i_{\rm s}=64.0^\circ$.

The stationary inclination for the disc is consistently slightly less than the value predicted for the ring in Fig.~\ref{fixedpoint}. However, the angular momentum of the disc and the binary evolve in the disc simulations. 
While the ratio oscillates because the angular momentum of the binary oscillates, the angular momentum of the disc generally decreases in time.

Furthermore, as described in the Introduction, the dynamics
of a ring are somewhat different from that of an extended disc.

\subsubsection{Condition for polar evolution}

For the simulations described in Section~\ref{eb0p5} for binary eccentricity $e_{\rm b}=0.5$ and a high mass disc $M_{\rm d}=0.05\,M$, the transition from librating to circulating solutions is in the range $50^\circ-60^\circ$. This system has angular momentum ratio $J_{\rm d}/J_{\rm b}=0.42$. The critical inclination between librating and circulating solutions for this high disc angular momentum in the analytic model is $51^\circ$ using Equation~(\ref{minlib2}).  For the same disc parameters except disc mass $0.001\,M$, we previously found that the critical angle was in the range $40^\circ-50^\circ$ \citep{Martin2018}. This system has angular momentum ratio $J_{\rm d}/J_{\rm b}=0.0084$. In the analytic model for low angular momentum ratio, the critical angle for this angular momentum ratio is $38.4^\circ$ (Equation~(\ref{minlib2})). The analytic model slightly underestimates the critical angle.

\section{Discussion}
\label{discussion}

The polar aligned disc observed in HD 98800  by \cite{Kennedy2019} is within $4^\circ$ of being perpendicular to the binary orbital plane.  The binary has a  semi-major axis of $a_{\rm b} = 1\,\rm au$, eccentricity 
$e_{\rm b} \simeq 0.785$, and the circumbinary gas disc in carbon monoxide  extends from about $1.6\,\rm au$ out to about $6.4\,\rm au$. Note that polar aligned discs have a smaller inner truncation radius than a disc aligned to the binary orbital plane \citep{Franchini2019b}. The binary component masses are
$0.699  M_\odot$ and $0.582  M_\odot$ \citep{Boden2005}. We assume that the disc has evolved to a stationary
configuration with tilt $i_{\rm s}$ given by Equation (\ref{farz}).
In making this assumption, we are assuming that this equation holds for a  radially wide disc with viscosity.

Using the Equation (\ref{farz}), we obtain an analytic expression for the ring (or disc)  to binary angular momentum ratio  
$j=J_{\rm r}/J_{\rm b}$, 
\begin{equation}
 j = \frac{( 1+ 4e_{\rm b}^2) \cos{i_{\rm s}}  } {3 (1-e_{\rm b}^2)  - 5 \cos^2{i_{\rm s}} }. 
\end{equation}
We apply the lower limit to the tilt $i_{\rm s}=86^{\circ}$ and $e_{\rm b} = 0.785$ to the above and
obtain
\begin{equation}
J_{\rm r} \simeq 0.21 J_{\rm b}.
\label{jvn}
\end{equation}

 Assuming the disc density falls off inversely with radius as $R^{-q}$ and using the disc inner and outer radii values, we have that 
 \begin{equation}
 J_{\rm r} = k M_{\rm r} \, a_{\rm b}^2 \Omega_{\rm b},
 \label{jvn1}
 \end{equation}
 where $\Omega_{\rm b}$ is the binary orbital frequency, and $M_{\rm r}$ is the mass of the ring. Quantity $k$ varies from 2.1 to 1.9 as $q$ varies from 0 to 1.5. 
Using the properties of the binary cited above, we have that 
\begin{equation}
J_{\rm b}= 0.15 M a_{\rm b}^2 \Omega_{\rm b}.
\label{jbn}
\end{equation}
Combining Equations (\ref{jvn}), (\ref{jvn1}), and (\ref{jbn}),
we have that $M_{\rm r} \simeq 0.016 M$. The disc (or ring) mass could be smaller if the
tilt is less than $4^\circ$ from perpendicular. Therefore,
the disc mass must be $\lesssim  0.016 M \approx 0.021\,\rm M_\odot$,  since the mass of the binary is $1.28\,\rm M_\odot$. This mass range is quite plausible
for protostellar discs.

Solid bodies may form within a gaseous disc that reaches its stationary inclination  $i_{\rm s}< 90^\circ$.
Such bodies will likely remain within the gaseous disc due to gravitational coupling, unless they are massive enough to open gaps.
As the gas disc dissipates, its tilt angle can increase until it reaches the polar state at $90^\circ$ misalignment with
respect to the binary. The orbits of the solid bodies will likely remain coplanar with the disc again due to gravitational coupling,
however, once the gaseous disc mass becomes sufficiently small this coupling will break down and the solid
bodies may decouple from the gas disc before it reaches its final value of $90^\circ$ inclination. Once the solid bodies break
free of the gas disc their libration speeds will no longer be coordinated and they will randomize relative to each other. 
The random velocities could affect the planet formation process. Just how this
operates is beyond the scope of this paper.

\section{Conclusions}
\label{conc}

In this work we have investigated the conditions under which the nodal libration mechanism can operate in a  protostellar disc around an eccentric binary as first described by \cite{Martin2017}. We apply both SPH simulations and analytic methods.  Such discs undergo oscillations of the tilt and longitude of ascending node, similar to test particle orbits. However, for the case of a disc, dissipation leads to polar alignment of the disc. We have investigated the effect of a nonzero mass disc on the system evolution. The mass of the disc affects the outcome of the process because the binary evolution is affected. The disc affects the binary orbit gravitationally and through advection of mass and angular momentum.
The binary eccentricity and tilt oscillate. The eventual alignment of  disc with nonzero mass is at an angle less  than $90^\circ$. This has significant implications for planet formation around eccentric binaries and for the detection properties of such discs.
 
We applied the secular evolution equations of \cite{Farago2010} to determine conditions related to the polar alignment of an arbitrary mass ring that orbits around
around an eccentric orbit binary.  We determined the stationary misalignment angle, the generalised polar angle,  between the ring and binary as a function
of system parameters. In the presence of dissipation, the ring tilt could evolve to this angle.
This angle, given analytically by Equations (\ref{farz}) and (\ref{isa}), decreases montonically
with increasing ratio of ring-to-binary angular momentum and decreasing binary eccentricity (see Figure~\ref{fixedpoint}). A very small mass ring lies perpendicular
to the binary orbit plane in the stationary configuration.

 We applied the stationary tilt angle Equation (\ref{farz}) to constrain the mass of the circumbinary disc in HD 98800 (see Section \ref{discussion}). We note that this condition is based on gravitational torques only and ignores the accretional torque on to the binary. Furthermore, it models the disc as a narrow ring.
 We did find however that SPH simulations appear to be in good agreement with the predictions of
Equation (\ref{farz})  (Section \ref{comp1}). In any case, we found this equation implies that the disc mass is less than about $0.02 M_\odot$, in the range of typical protostellar disc masses. More accurate observational determinations of the tilt angle would be of benefit. 

We determined analytic criteria required for a ring with mass to evolve to a generalised polar configuration (see Section \ref{sec:anasum}) and determined the minimum misalignment
inclination angles (see Figure~\ref{fig:CritINumerical}).  For small values of the disc-to-binary angular momentum ratio $j_0$, the minimum tilt angle increases with
with $j_0$. But for larger $j_0$, this angle decreases. The change in behaviour is understood in terms of a transition between different types of stationary points for marginally
librating orbits.
As discussed in Section \ref{compsa}, we found approximate agreement between the results of the SPH simulations and the analytic model.

\section*{Acknowledgments} 
 We thank Daniel
Price for providing the {\sc phantom} code for SPH simulations and
acknowledge the use of SPLASH \citep{Price2007} for the rendering of
the figures. SHL acknowledges useful discussions with Gordon Ogilvie. We acknowledge support from NASA through grants NNX17AB96G and 80NSSC19K0443.  Computer support was provided by UNLV's National
Supercomputing Center.

\bibliographystyle{mnras} 
\bibliography{ms}

\appendix
\section{Stationary Tilt}
We derive the conditions for the stationary tilt of the ring relative to the binary given in Equations(\ref{farx}) - (\ref{farz}) of Section \ref{sec:stationary}.
We apply the stationary  (fixed point) conditions of zero time derivatives (Equations (\ref{statcond1}) and (\ref{statcond2})) 
to the tilt and binary eccentricity evolution Equations (\ref{lx}) - (\ref{eb}). In addition, we apply the condition that $\ell_y=0$, since we are interested in tilts
in the $x$--$z$ plane. Equations (\ref{lx}),
(\ref{lz}), and (\ref{eb}) are trivially satisfied for zero time derivatives. Only Equation (\ref{ly}) for $d \ell_y/d \tau$ needs to be considered.
This equation implies
\begin{equation}
    \ell_x \left[(1+ 4 e_{\rm b}^2) \ell_z   
    + \frac{ \gamma_{\rm r}  }{\sqrt{1-e_{\rm b}^2}} \left( (1-e_{\rm b}^2)  (2- 5 \ell_x^2) + 5 e_{\rm b}^2 \ell_z^2 \right) \right]=0. \label{lyA}
\end{equation}
The solution with $\ell_x=0$ corresponds
to a coplanar system with the ring rotating
either prograde ($\ell_z=1$) or retrograde
($\ell_z=-1$) relative to the binary.

For  $\ell_x$ nonzero, the bracketed term in Equation (\ref{lyA}) is zero, resulting in the same equation as equation (A16)
of Appendix A4 in \cite{Farago2010} with some obvious changes in variable names.
We use the fact that $|\bm{\ell}|= \ell=1$ to  eliminate $\ell_x$ using $\ell_x^2=1-\ell_z^2$.
Equation (\ref{lyA}) can then be solved analytically as a
quadratic equation in $\ell_z$ to obtain two roots
\begin{equation}
    \ell_{z} =  \frac{ -(1+ 4e_{\rm b}^2)\pm \sqrt{ \left(1+ 4e_{\rm b}^2 \right)^2+60 (1-e_{\rm b}^2) j^2} }{10 j}.
\end{equation}

The solution with the positive sign for the square root has the property that $\ell_z$ goes
to zero as $j$ goes to zero. That is, a low mass ring is nearly perpendicular to the binary orbital plane. This solution is of interest for the purposes of this paper and is used
in the text as Equation (\ref{farz}).

The solution with the minus sign for the square
root applies to retrograde rings, since $\ell_z <0$. The requirement that $\ell_z > -1$
implies that
\begin{equation}
    j > \frac{1+4 e_{\rm b}^2}{2+3 e_{\rm b}^2}.
\end{equation}
In the limit of large $j$, we have that
\begin{equation}
\ell_z = - \sqrt{\frac{3 (1-e_{\rm b}^2)}{5}}.
\label{lzrjl}
\end{equation}
With increasing $j$, tilt component $\ell_z$ increases monotonically from $-1$ to the value given by Equation (\ref{lzrjl}).
We do not consider the applications of this stationary solution
in this paper.

\label{lastpage}
\end{document}